\documentstyle[multicol,prl,aps]{revtex}
\input epsf
\begin{document} 
\twocolumn
\title{
Electromagnetic Response of a Pinned Wigner Crystal
}
\author{H.A. Fertig}

\address{
Department of Physics and Astronomy, University of Kentucky,
Lexington, Kentucky 40506-0055.
}

\address{\mbox{ }}

\address{\parbox{14.5cm}{\rm \mbox{ }\mbox{ }
A microscopic model for analyzing the microwave absorption properties of
a pinned, two-dimensional Wigner crystal in a strong perpendicular magnetic
field is developed.  The method focuses on excitations within the lowest
Landau level, and corresponds to a quantum version of the harmonic
approximation.  For pure systems (no disorder), the method reproduces
known results for the collective mode density of states of this system,
and clearly identifies the origin of previously unexplained structure
in this quantity.  The application of the method to a simple diagonal
disorder model uncovers a surprising result: a sharp (delta-function)
response at zero temperature that is consistent with recent experiments.
A simple spin lattice model is developed that reproduces
the results of the quantum harmonic approximation, 
and shows that the sharp response is possible because the
size scale $L_c$ of patches moving together in the lowest frequency
collective mode is extremely large compared to the
sample size for physically relevant parameters.  This result
is found to be a direct repercussion of the long-range nature
of the Coulomb interaction.
Finally, the model is used to analyze
different disorder potentials that may pin the Wigner crystal, and it is
argued that interface disorder is likely to represent the dominant
pinning source for the system.  A simple model of the interface
is shown to reproduce some of the experimental trends for the magnetic field 
dependence of the pinning resonance.
}}
\address{\mbox{ }}
\address{\mbox{ }}
\address{\parbox{14.5cm}{\rm PACS numbers: 73.40.Hm, 73.20.Mf, 75.40.Gb }}
\maketitle

\section{Introduction}

It has long been appreciated that the groundstate of electrons in
an otherwise structureless environment should be crystalline at
low enough densities\cite{wigner}.  Considerable effort has been
focused on creating such a state in a two-dimensional
electron gas (2DEG) as realized in semiconductor 
heterojunction and quantum well
systems\cite{review}, although obtaining the appropriate
limit of low electron and impurity densities has proven difficult.
The introduction of a magnetic field perpendicular to the
2DEG improves this situation by raising the electron density
at which crystallization is thought to occur.  Recent
experiments have identified the onset of strong insulating
behavior in very high mobility systems, at magnetic fields
such that the filling factor $\nu=N/N_{\phi}$, with $N$ the
number of electrons and $N_{\phi}$ the number of magnetic
flux quanta through the system, is below $\sim1/5$ for 
electrons\cite{jiang90,goldman90}
and $\sim1/3$ for hole systems\cite{santos92}. 
While much intriguing experimental data has accumulated,
a definitive proof that the low filling factor insulating
state of these two-dimensional systems is indeed an
electron, or Wigner, crystal (WC) has remained elusive.

One experimental fact for the low filling factor systems
that is in clear agreement with expectations for a WC
interpretation is that they are insulating.  It is well
appreciated by now, from analogous behavior in charge 
density wave (CDW) systems\cite{gruner}, that an arbitrarily
small disorder potential should pin the WC at zero
temperature so that there are no charge carriers that
can flow in response to an arbitrarily small 
static electric field.
In CDW systems,
the pinning potential also supplies a restoring force
that induces a broad peak in the ac electromagnetic response
at the ``pinning frequency'', 
whose magnitude may be used to assess the correlation
length of the CDW\cite{gruner,fukuyama78}.  Early experiments
on the magnetically-induced WC identified similar structure
in the density response using various 
techniques\cite{andrei88,paalanen92,willett94,li95}.  
Theoretical 
analyses\cite{glattli90,normand92,ferconi93} of these 
experiments have largely focused on determining
the correlation length of the WC from the experimental
data, although no clear consensus on this quantity has 
yet been reached.

Very recently, experiments on high quality hole\cite{li97,mellor97}
and electron\cite{engel97} systems at low filling
factors have revealed structure
in their microwave absorption properties that
are {\it qualitatively different} than what is observed
in CDW systems.  At the lowest temperatures and highest
magnetic fields available, these systems exhibit {\it sharp}
resonances at low frequencies, with quality factors 
$Q=f/\Delta f$ ($f$ here is the frequency of the resonance
peak and $\Delta f$ its width) as high as 30.
The frequency of the peak increases with
increasing magnetic field, and may saturate at
a maximum for the largest fields\cite{li97}.
By contrast, most existing theoretical work
predicts\cite{glattli90,normand92,fukuyama78II,ferccom} 
a broad ($Q \sim 1$)
resonance, and a pinning frequency that {\it decreases}
with magnetic field.  The subject of this work is
to understand some aspects of this puzzling experimental
finding.

In what follows we will adopt a microscopic
model of the magnetically induced two-dimensional
Wigner crystal.  The groundstate is 
assumed to be well described by a product of 
localized Gaussian wavepackets, so that exchange
effects are ignored\cite{maki83}.  Our goal is
to find the response of the system to a time dependent,
spatially uniform electric field\cite{kohn61}.  
To compute the latter
quantity we will employ a ``quantum harmonic approximation'' 
(QHA)\cite{fredkin},
a natural generalization of the classical harmonic
approximation, in which a finite number of angular
momentum orbitals per site is retained in the Hilbert
space of states for the electron system.  This is of
course sensible provided that the electrons remain
close to their groundstate sites after being
excited by the electric field.  A very strong
magnetic field is assumed so that projection into
the lowest Landau level is appropriate\cite{com0}, and all
results reported here are for zero temperature.
For an undisordered WC, the QHA reproduces known
results found {\it without} resorting to a
harmonic approximation\cite{cote91}, and indeed
offers an explanation of features found in
the density response function that were not
easily interpreted previously.

Because exchange effects are negligible in this system\cite{maki83},
a time-dependent Hartree approximation turns out to
be both convenient and accurate for computing the
response functions of this system.  For simplicity
we consider only a diagonal disorder model; i.e.,
the disorder enters only in the single particle
excitation energies and not in the interaction
matrix elements among the electrons.  
In practice this means we have adopted
a model in which the lattice is perfectly
ordered, and disorder is introduced as
a random on-site pinning potential\cite{merkt96}.
As we will see, this
assumption greatly simplifies the
computation of matrix elements entering the QHA.
Using a perturbative approach described below,
for weak disorder
it is possible to demonstrate that our
qualitative results 
are insensitive to the diagonal disorder assumption.

For all the disorder models studied in this work,
the results of the QHA turn out to be qualitatively
the same: the response functions exhibit an extremely
sharp response at the lowest collective mode
frequency (provided that the electron-electron
interaction is unscreened; see below.)
This is the central result of this work.
Two physical models of pinning are analyzed in detail in
this study.  Charged impurities which may be
found in the spacer layer between donors and
the electrons have been argued\cite{ruzin92}
to represent the strongest source of pinning
for the WC in dc non-linear $I-V$ measurements;
i.e., these set the energy scale necessary to 
fully dislodge the electrons in a non-vanishing static
electric field so that a current may flow. 
From the
CDW point of view\cite{fukuyama78}, this model
corresponds to a strongly pinned system.  However,
we will see that the strong pinning approach  
grossly overestimates the pinning frequency,
so that the
experimentally measured pinning frequencies cannot
be explained by this source of disorder\cite{com1}.
The ultimate reason for the discrepency between the
pinning frequency in the microscopic model and the CDW result
for this is that a magnetically-induced WC is {\it not}
a CDW.  Our results indicate that the collective
mode spectrum associated with strongly pinned
centers is far more like that of a crystal
with vacancies pinned at charged impurity sites than
that of an elastic medium tied down at random sites.

We then focus on a model
of interface disorder that we believe is likely
to be the dominant pinning mechanism in
electromagnetic absorption for these systems.
Heterostructure and quantum well interfaces
are believed to have structure at length scales
of several tens of angstroms\cite{interface}.
An interpretation of this is that in semiconductor
heterostructures the interface between different
materials (typically GaAs and AlAs) 
may only be defined
to within one or two lattice constants ($\sim 5\AA$).
This is often modeled as an interface with
pits and/or islands of typical size scale
$< 100\AA$ \cite{leo89}.  Electrons residing
in large pits or regions with an unusually
large number of small pits
have an enhanced probability
of lying slightly closer to the
donor layer and so may be bound by them.
An important aspect of the physics
in this model is that in the lowest Landau level
interfaces with 
pits whose size scale are smaller than the
magnetic length $l_0=(\hbar c / eB)^{1/2}$ 
have pinning potentials that increase
with decreasing $l_0$.
This leads to an {\it increasing} pinning
frequency with magnetic field, as seen in experiment.
The $\sim$1 GHz magnitude of the measured pinning frequency
may be explained with very reasonable parameters
describing the interface at which the 2DEG resides,
as described below.

A typical result
for a single disorder realization, whose precise form is described
below,
is shown in Fig. 1.  
The system includes 1024 electrons, and is 
assumed here (as throughout this work) to 
obey periodic boundary conditions.
Because of the finite number of degrees of freedom,
the density response function consists of a series
of delta functions.
Note that the scale 
of the figure is logarithmic, so that a linear plot
of power absorbed as a function of frequency shows
a single delta-function response at the lowest
collective mode frequency, as illustrated
in the inset.  The height of
this peak shows no sign of decreasing with increasing
system size, nor do any other collective modes
develop a noticeable oscillator strength in
power absorption.  For a model in which the typical
on-site pinning potential is small, we find
the pinning frequency (i.e., the position of the
delta-function response in Fig. 1) 
to a good approximation is given by
$v_0$, the energy required
to excite an electron from the $m=0$ to the $m=1$ angular
momentum state in the absence of other electrons,
averaged over all the electrons.
This value is is essentially the same as the
well-known result for weakly pinned CDW's\cite{fukuyama78}.
Typically, the exact pinning frequency found in the
QHA falls somewhat below this estimate, by an
amount that is specific to the precise disorder
realization.  This correction to the weak pinning
result may be estimated using a perturbative
approach described below.

Although the sharpness of the absorption peak seems to
be in agreement with recent experiments, one needs some
analytic method to demonstrate that in the thermodynamic
limit the peak does not broaden, particularly since
this result is so different than prior 
expectations\cite{normand92,fukuyama78II}.
Towards this end we develop a (pseudo)spin lattice model
in which the $\hat{z}$ component of the spin at a given site
represents the angular momentum state of an electron.
A convenient perturbation theory
for the system may be developed around a uniformly
pinned state (i.e., one in which the pinning potential
for every site is $v_0$.)  The perturbing
parameter is then $\Delta U(\vec{R})$, with $\lbrace\vec{R}\rbrace$
representing the sites around which the electrons are
localized and $U(\vec{R})=v_0+\Delta U(\vec{R})$
the energy required to excite an electron from the
$m=0$ to the $m=1$ state.  
For the uniformly pinned system, it is natural for the power
absorbed to have a $\delta$-function response, since
the collective modes have wavevector $\vec{q}$
as a good quantum number.  The correction to the
weight $W$ of this $\delta$-function due to
disorder may be computed in perturbation theory,
and a depletion $D$ may defined such that
$W \propto 1-D$.  To lowest non-trivial order
in perturbation theory, the largest contribution
to $D$ is found to have
the form:

\begin{equation}
D \propto \int d^2 q {{|\Delta U(\vec{q})|^2 
|u_{\vec{q}}^+|^2 } \over
{|E^p_{\vec{q}} - E^p_{\vec{q=0}} |^2} }
\label{approx_overlap}
\end{equation}

Here $\vec{q}$ represents wavevector, $\Delta U(\vec{q})$
the Fourier transform of the perturbation, and
$E^p_{\vec{q}}$ the collective mode dispersion
for the uniformly pinned system.  
$|u_{\vec{q}}^+|^2$ is a weighting function
whose precise form is given below; in the
limit $q \rightarrow 0$, $|u_{\vec{q}}^+| \rightarrow 1$.
Provided that the disorder potential does not have
fluctuations on arbitrarily long length scales --
i.e., that arbitrarily large 
patches of unpinned or strongly pinned electrons are rare --
then one expects
$|\Delta U(\vec{q})|^2 \propto q^2$ for small $q$.
The dispersion relation for the uniformly pinned
system we show below has the form
\begin{equation}
E^p_{\vec{q}} \approx v_0 + 2\pi e^2 \rho_0 l_o^2 q /\kappa
+ O(q^2)
\label{dispersion}
\end{equation}
for small $q$, where $\rho_0$ is the electron density
and $\kappa$ is the dielectric constant of the host
semiconductor.  
The linear dispersion with $q$ turns out to be a direct
result of the long-range nature of the Coulomb interaction,
and is not present in a model with short-range (e.g., screened) 
electron-electron interactions.  Plugging Eq. \ref{dispersion}
into Eq. \ref{approx_overlap} demonstrates that 
for a given disorder realization
a finite depletion
results that will be small if the pinning is not too far
from uniform.  This directly demonstrates that
it is the long-range nature of the electron-electron
interaction that is responsible for the sharp response in the system.
A system with short-range interactions has a collective
mode spectrum dispersing {\it quadratically} with $q$ away
from $v_0$, and as seen for Eq. \ref{approx_overlap}, this
leads to a divergence in $D$, signaling that
the response is {\it not} sharp in this case.  Fig. 2
illustrates the response function for a model in which
the electron-electron interaction is screened.  As may be
seen, even for a relatively small number of particles ($N=225$),
the response has been significantly broadened, as is
typically found in CDW systems\cite{gruner,fukuyama78,normand92}.
Roughly speaking, this
says that an appropriate analogy for the randomly pinned
WC in a strong magnetic field is a hard object that does
not deform much when vibrating on randomly placed springs,
so that a well-defined periodic response is to be 
expected in spite of the randomness of the pinning.

One important caveat to this result must be noted.
The assumption that $|\Delta U(\vec{q})|^2 \propto q^2$
is actually not true for the most common forms of
disorder studied. Specifically, white noise pinning
models usually have equal fluctuation strengths at all
length scales, so that $\Delta U$ does not vanish
as $q \rightarrow 0$.  
In our expression Eq. \ref{approx_overlap},
this leads to a formal logarithmic divergence 
in the thermodynamic limit and an expected
broadening of the absorption peak.  We note however that
the disorder models used in the QHA have a white-noise
character, yet we have not been able to detect a finite width
in the resonance for any disorder realization.  
Presumably this indicates that the $\delta$-function
response is a finite system size effect, and one needs
to estimate a length scale above which broadening in
the response function should become apparent.  This
may be accomplished by noting that the integral
in Eq. \ref{approx_overlap} has an infrared cutoff
$2\pi/L$ where $L$ is the system size.  The value of
$L$ for which $D \approx 1$, which we call $L_c$,
sets the size scale above which broadening will
be significant.  Solving for $L_c$, one finds
$$
L_c \approx \ell\exp
\bigl\lbrace
{{\nu^3} \over {|\Delta \tilde{U}|^2} }
 \bigr\rbrace ,
$$
where $\ell$ is a length scale above which Eq. \ref{dispersion}
is accurate [$\ell \approx 10a_0$, with $a_0$ the WC
lattice constant], and $\Delta \tilde{U} =
\lim_{q \rightarrow 0} \Delta U(\vec{q}) \kappa l_0/e^2$.

The physical interpretation of $L_c$ is that it is the
length scale for coherent motion of patches of electrons
in the lowest collective mode, since for $L \ll L_c$ the
perturbation theory is valid and the system responds much like
as a single oscillator to the excitation field.  A conservative
estimate of $L_c$ for the experiments of 
Refs. \onlinecite{li97,mellor97} shows that in practice
it is huge, $L_c > 10^{24}a_0$ at $\nu=0.22$,
where $a_0$ is the lattice constant of the WC.
Thus, remarkably, in spite of the formal divergence
in the thermodynamic limit,
the perturbation theory is in fact controlled and valid
for samples with physically relevant dimensions,
where $L \sim 10^5a_0 \ll L_c$.
Thus, our interpretation of the WC response as 
one of an undeformable oscillator is indeed appropriate.

This article is organized as follows.  In Section II, we develop
the QHA used to compute the electromagnetic response of the
pinned WC.  Section III develops the pseudospin approach
to the collective modes and shows how a perturbation theory
may be developed around the uniformly pinned state,
and the calculation of the depletion $D$ is discussed.
Section IV discusses some details of the interface
pinning model.
Readers interested in our results and not the details
of the calculations may wish to proceed directly to
Section V, 
which discusses results for different pinning models,
including
charged impurities in the electron layer and the interface
pinning model.  Section VI discusses the relationship of this
work with results of other calculations, 
as well as several unresolved questions regarding 
the experimental results.
We conclude with
a summary in Section VII.

\section{Quantum Harmonic Approximation}
\label{sec:qha}

In this section we will develop a quantum mechanical generalization
of the harmonic approximation, in which the electrons are
treated as distinquishable, so that the groundstate and
low-lying excited states may be represented as linear combinations
of (unsymmetrized)
products of single particle states.
The single particle states we will use consist of
angular momentum states, and we will find that
it is an excellent approximation to retain only
the lowest lying ones 
when computing low-energy properties
of the system.  Once we have truncated the
Hilbert space in this way, it becomes possible
to numerically compute response functions for reasonably
large systems, which
one may compare with experiment.

The groundstate of two-dimensional charged particles in a very
strong magnetic field is believed\cite{maki83}
to be accurately represented by a collection 
of distinquishable particles in Gaussian
orbitals of the form
$$
\phi_{0,\vec{R}_i}(\vec{r}_i)=
\bigl({1 \over {2\pi l_0^2}}\bigr)^{1/2}
e^{-|\vec{r}_i-\vec{R}_i|^2/4l_0^2+ {i \over 2}
\hat{z}\cdot(\vec{R}_i \times \vec{r}_i)/l_0^2}.
$$
In this state, the electron with coordinate $\vec{r}_i$
is maximally localized within the lowest Landau level.
Since the kinetic energy of a collection of electrons
in such orbitals is already as small as possible,
the groundstate presumably will be found by minimizing
the total potential energy with respect to the parameters
$\vec{R}_i$. 
In the absence of disorder these should
be chosen to lie on a perfect lattice, so that the charge
distribution in the limit of infinite magnetic field
($l_0 \rightarrow 0$) approaches that of 
the groundstate for a distribution
of {\it classical} point particles.  
Disorder changes the optimal positions
for the electrons\cite{cha94}, but provided the
disorder is not too strong, the site centers
$\vec{R}_i$ will presumably not be close to one
another on the scale of the magnetic length $l_0$.
In this situation, wavefunction overlaps among the 
individual electrons are negligible, so that a product
of the single particle wavepackets $\phi_{0,\vec{R}_i}(\vec{r}_i)$
is an accurate representation of the groundstate, despite
the fact that it is not explicitly antisymmetric\cite{maki83}.

Since we are interested
in the electromagnetic response of the system at low frequencies,
it will be necessary to compute the low-lying excited states
of the system.  In a classical analysis, this is typically
done by assuming the displacements of the electrons from their
groundstate positions are small, so that the energy may be
expanded to second order in displacements, allowing the normal
modes of the system to be found in a relatively simple fashion.
If we assume the electrons are localized within
a magnetic length of their
groundstate positions, then the analog of the small displacements 
approximation is to allow individual electrons to be excited
into higher angular momentum states localized about each lattice
site.  Provided the angular momentum is not too large, the
charge densities associated with these states 
for a given site will not overlap
significantly with those of its neighbors, so that exchange
effects may continue to be ignored.  Thus we consider a set
of states for each site
\begin{eqnarray}
\label{orbitals}
\phi_{m,i}(\vec{r}) &=&
\bigl({1 \over {2\pi l_0^2 2^m m!}}\bigr)^{1/2}
\bigl({{z - Z_i} \over {l_0}}\bigr)^m \nonumber \\
&\times&
e^{-|\vec{r}-\vec{R}_i|^2/4l_0^2+ {i \over 2}
\hat{z}\cdot(\vec{R}_i \times \vec{r})/l_0^2},
\end{eqnarray}
where $z=x+iy$ is the electron position in complex notation,
and $Z_i=R^x_i+iR^y_i$.
If we define creation operators for these single particle
states $a_{mi}^{\dag}$, then the Hamiltonian for the
system may approximately be written in the form
$$
H=\sum_{i}\sum_{mn} V^{i}_{mn} a^{\dag}_{mi} a_{ni}
$$
$$
+{1 \over 2} \sum_{i \ne j} \sum_{m_i n_i m_j n_j}
U_{m_i n_i m_j n_j}^{ij} 
a^{\dag}_{m_i i} a_{n_i i} a^{\dag}_{m_j j} a_{n_j j}.
$$
Here, $V^{i}_{mn}$ represents the interaction of
an isolated electron with a disorder potential, and
the interaction matrix element is given by
\begin{eqnarray}
\label{matrix_element}
U^{ij}_{m_1 m_2 m_3 m_4} =
\int d^2r_1 d^2 r_2 \bigl({ 1 \over {2\pi l_0^2} }\bigr)^2
\bigl[ { 1 \over { 2 l_0^2 } } \bigr] ^ {(m_1+m_2+m_3+m_4)/2}
\nonumber \\ \times
z_1^{*m_1}z_1^{m_2} z_2^{*m_3} z_2^{m_4}
e^{-r_1^2/2l_0^2-r_2^2/2l_0^2}
v(\vec{r}_1-\vec{r_2}+\vec{R}_{ij}),
\end{eqnarray}
where for most cases we will study, $v(r)$ is the Coulomb
interaction $e^2/\kappa r$, with $\kappa$ the dielectric
constant of the host material for the electron layer.

To simplify our calculations, we will make certain assumptions
about the form of the disorder entering $H$.  Firstly,
we assume that although disorder certainly will cause
the orbit centers $\vec{R}_i$ to vary from the positions
of a perfect lattice, this variation does not affect the
qualitative features of the absorption spectrum.  
Thus in practice our $\vec{R}_i$'s take on the
values of perfect triangular lattice positions.  It should
be noted that this choice of the $\vec{R}_i$'s is not required
to carry out the QHA, but by adopting it the numerical
computation of $U_{m_i n_i m_j n_j}^{ij}$ is greatly
simplified.  Some details on how this is done
in practice are discussed in Appendix A.
The second simplification we introduce is to
assume that the pinning potential at the individual sites
is circularly symmetric, so that in the groundstate
the electrons occupy the $\phi_{m=0,i}$ orbitals and do not
admix higher angular momentum states.  Again, the
QHA may be developed without this assumption.  However,
this simplification has the great advantage of allowing
an analytic specification of the groundstate.  In the
absence of this assumption one would need to find the
groundstate orbital occupations numerically, which 
presumably could be computed with sufficient accuracy
using a static Hartree approximation.  While such
a calculation is clearly feasible, we believe that
allowing assymmetric forms for $V^{i}_{mn}$ will 
have little quantitative effect, provided the average
of this quantity over the sites restores the circular
symmetry.  Finally, to take full advantage of the
symmetries of the system, we will impose periodic 
boundary conditions.

A second caviat related to this is that in principle
one must retain fewer than seven orbitals on each site
to have each electron lie purely in the $m=0$ orbital
in the groundstate.  Even in the absence of disorder,
the six-fold symmetry of the lattice will in principle
admix in states with angular momenta equal to integral
multiples of six.  Although this effect is extremely
small\cite{maki83}, it can in principle lead to weak
instabilities in the response function computed below
when too many orbitals are included per site.  As we
will see, all the low energy excitations (i.e., phonons)
of the system are accurately captured when only {\it two}
orbitals per site are retained, so that in practice this
does not arise as a problem.

With these simplifications, the Hamiltonian takes the form
\begin{eqnarray}
H=\sum_{i}\sum_{m} \varepsilon_{i}(m) a^{\dag}_{mi} a_{mi}
\quad\quad\quad\quad
\nonumber \\
+{1 \over 2} \sum_{i \ne j} \sum_{m_i n_i m_j n_j}
U_{m_i n_i m_j n_j}^{ij} 
a^{\dag}_{m_i i} a_{n_i i} a^{\dag}_{m_j j} a_{n_j j}.
\label{hamiltonian}
\end{eqnarray}
The zero of energy may always be chosen such that
$\varepsilon_{i}(m=0)=0$, and we expect $\varepsilon_{i}(m>0) \ge 0$.
The choice of $\lbrace\varepsilon_{i}(m)\rbrace$ defines
the specific disorder model we are studying.  We note
that the greatest power of this method is that it
is well-suited to disorder potentials that vary on length
scales of the order of $l_0$ and smaller.  The two
pinning mechanisms we study in detail -- charged impurities
and interface pinning -- both fall into this category.

The quantity we will calculate, from which either power absorption
or frequency dependent conductivity may be computed, is the
response function
\begin{equation}
\chi_{m_1m_2m_3m_4}(ij;\tau)=
-<T_{\tau} a^{\dag}_{m_1i}(\tau) a_{m_2i}(\tau) 
a^{\dag}_{m_3j}(0) a_{m_4i}(0)>
\label{response_function}
\end{equation}
where here $\tau$ is imaginary time and $a_{mi}(\tau),~a_{mi}^{\dag}(\tau)$
are the usual time-dependent Heisenberg representations of the
annihilation and creation operators, and $T_{\tau}$
the time ordering operator\cite{mahan}. The brackets
$<\cdot\cdot\cdot>$ represents a thermal average.
The Fourier
transform of this function with respect to imaginary time
has poles at the collective mode frequencies.  To generate
a closed formula for $\chi$, we employ an equation of motion
method similar to that used in Ref. \onlinecite{cote91}.
The time derivative of $\chi$ satisfies the equation
$$
{{\partial \chi_{m_1m_2m_3m_4}(ij;\tau)} \over {\partial \tau} }
=-<[\rho_{m_1m_2}(i;\tau), \rho_{m_3m_4}(j;0)]>\delta(\tau)
$$
\begin{equation}
-<T_{\tau} {{\partial \rho_{m_1m_2}(i;\tau)} \over {\partial \tau} }
\rho_{m_3m_4}(j;0)>
\label{time_deriv_chi}
\end{equation}
where $\rho_{m_1m_2}(i;\tau)=a^{\dag}_{m_1i}(\tau) a_{m_2i}(\tau)$.
The time derivative of $\rho$ may be computed from the 
Heisenberg equation of motion
$$
{{\partial \rho_{m_1m_2}(i;\tau)} \over {\partial \tau} }
=[H-\mu N, \rho_{m_1m_2}(i;\tau)].
$$
Note that the formal inclusion of the chemical 
potential $\mu$ is necessary because
of the use of the finite temperature formalism; this means
our formalism allows in principle for us to treat the
situation in which there is more than one electron
per site.  In what follows, $\mu$ will be chosen such
that there is one electron per site in the groundstate,
and we will see in any case that any formal dependence
on $\mu$ drops out of our equations for the response
functions, so that it does not need to be
explicitly computed.  The relevant commutators in the equation
of motion may be worked out using 
$\lbrace a_{mi},a_{nj}^{\dag} \rbrace= \delta_{mn}\delta_{ij}$
\cite{com2}
from which one may show 
\begin{equation}
[\rho_{m_1m_2}(i),\rho_{m_3m_4}(j)]=
\bigl( \rho_{m_1m_4}(i) \delta_{m_2m_3} - 
\rho_{m_3m_2}(i) \delta_{m_1m_4} \bigr) \delta_{ij}.
\label{commutator}
\end{equation}
Combining Eqs. \ref{hamiltonian}, \ref{time_deriv_chi}, and
\ref{commutator}, the equation of motion takes the form
$$
{{\partial \chi_{m_1m_2m_3m_4}(ij;\tau)} \over {\partial \tau} } =
\quad\quad\quad\quad
$$
$$
-\bigl( <\rho_{m_1m_4}(i)>\delta_{m_2m_3} - 
         <\rho_{m_3m_2}(i)>\delta_{m_1m_4} \bigr) \delta_{ij}\delta(\tau)
$$
$$         
+ \bigl( \varepsilon_i(m_1) - \varepsilon_i(m_2) \bigr) 
   \chi_{m_1m_2m_3m_4}(ij;\tau)
$$
$$
-\sum_{n_1n_2n_3n_4} \sum_l U^{\prime~il}_{n_1n_2n_3n_4}
<T_{\tau} \bigl[ \rho_{n_1m_2}(i;\tau) \rho_{n_3n_4}(l;\tau) \delta_{n_2m_1}
$$
\begin{equation}
       - \rho_{m_1n_2}(i;\tau) \rho_{n_3n_4}(l;\tau) \delta_{n_1m_2} \bigr]       
       \rho_{m_3m_4}(j;0) >.
\label{exact_eq_mot}
\end{equation}
Here $U^{\prime~il}_{n_1n_2n_3n_4} = U^{il}_{n_1n_2n_3n_4}(1-\delta_{il})$.
To obtain a closed form for the equation of motion, we utilize a Hartree
decomposition of the last term in Eq. \ref{exact_eq_mot}.  This is
very much in the spirit of Ref. \onlinecite{cote91}; however, because
we are ignoring overlap among single particle states located at
different sites, it is unnecessary (and in fact would be inappropriate)
to include exchange terms in the decomposition.  We thus make the
substitution
$$
\bigl( \rho_{n_1m_2}(i;\tau) \rho_{n_3n_4}(l;\tau) \delta_{n_2m_1}
       - \rho_{m_1n_2}(i;\tau) \rho_{n_3n_4}(l;\tau) \delta_{n_1m_2} \bigr)       
       \rightarrow
$$
$$       
\bigl( \rho_{n_1m_2}(i;\tau)\delta_{n_2m_1} - 
       \rho_{m_1n_2}(i;\tau)\delta_{n_1m_2} \bigr) <\rho_{n_3n_4}(l)>
$$
$$       
       +
\bigl( <\rho_{n_1m_2}(i)>\delta_{n_2m_1} - 
       <\rho_{m_1n_2}(i)>\delta_{n_1m_2} \bigr) \rho_{n_3n_4}(l;\tau).      
$$
Substituting the above decomposition and Fourier transforming 
Eq. \ref{exact_eq_mot} with respect to imaginary time, we arrive
at the time-dependent Hartree approximation for the response
function:
\vbox{
$$
i\omega_n \chi_{m_1m_2m_3m_4}(ij;i\omega_n) =
\quad\quad\quad\quad
$$
$$
-\bigl( <\rho_{m_1m_4}(i)>\delta_{m_2m_3} - 
         <\rho_{m_3m_2}(i)>\delta_{m_1m_4} \bigr) \delta_{ij}
$$
$$         
+ \bigl( \varepsilon_i(m_1) - \varepsilon_i(m_2) \bigr) 
   \chi_{m_1m_2m_3m_4}(ij;i\omega_n)
$$
$$
+\sum_{n_1n_2n_3n_4} \sum_l U^{\prime~il}_{n_1n_2n_3n_4}
\biggl\lbrace
\bigl[ \chi_{n_1m_2m_3m_4}(ij;i\omega_n)\delta_{n_2m_1}
$$
$$
    -\chi_{m_1n_2m_3m_4}(ij;i\omega_n)\delta_{n_1m_2} \bigr]<\rho_{n_3n_4}(l)>
$$
$$    
+ \bigl[<\rho_{n_1m_2}(i)>\delta_{n_2m_1}
          -<\rho_{m_1n_2}(i)>\delta_{n_1m_2} \bigr]
$$
\begin{equation}  
            \times \chi_{n_3n_4m_3m_4}(lj;i\omega_n) \biggr\rbrace.
\label{eq_motion}
\end{equation}
}  
Once we have solved Eq. \ref{eq_motion}, in principle we may
compute any response function we like.  However, since ultimately
we are interested in the conductivity or power absorption
of the system for a spatially uniform electric field, it
is convenient to formulate the equation for a response
function that is less cumbersome in terms of indices, but
which may still be used to compute the physical quantities
of interest.  Towards this end we just consider the center
of mass response of the system.  An operator corresponding
to the displacement of the
center of mass may be written in the form
\begin{equation}
\vec{u}_{CM}={1 \over N} \sum_{m_1m_2} \sum_i 
<m_1,i|\vec{r}-\vec{R}_i|m_2,i>
a_{m_1,i}^{\dag}a_{m_2,i},
\label{ucm}
\end{equation}
where $|m_2,i>$ is a ket-vector representation of $\phi_{m,i}(\vec{r})$.
The perturbation due to a time-dependent, spatially
uniform electric field (e.g., microwaves) in terms of
$\vec{u}_{CM}$ is $-eN\vec{E}_0 \cdot \vec{u}_{CM}$,
where $N$ is the total number of electrons.  A 
quantity whose response to this perturbation is convenient
to study is the center of mass displacement itself, so
that the response function we will actually focus on is
\begin{equation}
\chi^{\alpha\beta}_{CM}(\tau)=-<T_{\tau} u_{CM}^{\alpha}(\tau)
u_{CM}^{\beta}(0)>,
\label{cmresponse}
\end{equation}
where $\alpha,\beta = x,y$.

The conductivity and hence power absorption may be written
in terms of this quantity as follows.  The time Fourier
transform of the spatially averaged current density
in response to an applied external field  
${\rm Re}\vec{E}_0e^{i\omega t}$ is
$<\vec{j}(\omega)> = ie\rho_0\omega \vec{u}_{CM}(\omega)$,
with $\rho_0$ the sheet density of the electrons.
In linear response theory, this takes the form
(at zero temperature)
$<\vec{j}(\omega)> = -ie^2\rho_0\omega N \sum_{\beta}
\chi^{\alpha\beta}_{CM}(\omega+i\delta) E_0^{\beta}$, where in the
usual way\cite{mahan} we have Fourier transformed the
response function with respect to imaginary time, and
made the replacement $i\omega_n \rightarrow \omega+i\delta$
to take the zero temperature limit.
The dot product of this quantity with the total electric
field is proportional to the power absorbed.  A minor
complication is that one must include the
screening field generated by the displacements of the
electrons.  Since we are interested in the
bulk current through the system, we assume that the induced
electric field may be replaced by its spatial average\cite{com3}.
Using a dipole approximation for the electric field generated
by the motion of the electrons, it is easily shown that
$\vec{E}_{ind}(t) = {e \over 2} \alpha \vec{u}_{CM}(t)$,
with $\alpha=\sum_{i\ne 0}1/R_i^3$.  Using $\vec{E}_{tot}
=\vec{E}_{0} + \vec{E}_{ind}$, one finds
for the power absorption 
$<\vec{j}(\omega)> \cdot \vec{E}_{tot} \propto 
\sum_{\alpha,\beta} E^{\alpha}_{tot} \sigma_{\alpha\beta} (\omega) 
E^{\beta}_{tot}$, with a the conductivity matrix given by
$$
{ \sigma}(\omega) = -ie^2\rho_0\omega N
{ \chi}_{CM}(\omega+i\delta)[{\bf 1} - 2\alpha N
{ \chi}_{CM}(\omega+i\delta)]^{-1}
$$
where here we are regarding ${ \sigma}$ and
${ \chi}_{CM}$ as 2 $\times$ 2 matrices.
As has been pointed out before\cite{cote92}, the
induced electric field shifts the frequency of
the peaks in ${\bf \sigma}(\omega)$ from where they
are found in ${\bf \chi}_{CM}(\omega+i\delta)$.
However, since the width of the  
peak we will find in ${\bf \chi}_{CM}$
is remarkably small, this shift may be neglected
for our purposes.  Thus in this study we will
focus on ${\bf \chi}_{CM}$ and its frequency dependence.

To compute the center of mass response, we define
an intermediate response function
$$
\chi^{\alpha}_{m_1m_2}(i;\tau) \equiv
\sum_j\sum_{m_3m_4} <m_3,j|r^{\alpha}-R^{\alpha}_j|m_4,j>
$$
$$
\times
\chi_{m_1m_2m_3m_4}(ij;\tau)
$$
in terms of which 
$$\chi^{\alpha\beta}_{CM}(\omega)={1 \over {N^2}}
\sum_i\sum_{m_1m_2} <m_1,i|r^{\alpha}-R^{\alpha}_i|m_2,i>
\chi^{\beta}_{m_1m_2}(i;\omega).$$  
Combining the definition
of
$
\chi^{\beta}_{m_1m_2}(i;\tau)
$ with Eq. \ref{eq_motion}, one derives the equation of motion
\vbox{
$$
i\omega_n \chi_{m_1m_2}^{\beta}(i;i\omega_n) =
\quad\quad\quad\quad
$$
$$
\sum_m \bigl[ <\rho_{mm_2}(i)><m|r^{\beta}|m_1> 
$$
$$
\quad\quad\quad\quad 
        - <\rho_{m_1m}(i)><m_2|r^{\beta}|m> \bigr] 
$$
$$         
+ \bigl[ \varepsilon_i(m_1) - \varepsilon_i(m_2) \bigr] 
   \chi_{m_1m_2}^{\beta}(i;i\omega_n)
$$
$$
+\sum_{n_1n_2n_3n_4} \sum_l U^{\prime~il}_{n_1n_2n_3n_4}
\biggl\lbrace
\bigl[ \chi_{n_1m_2}^{\beta}(i;i\omega_n)\delta_{n_2m_1}
$$
$$
    -\chi_{m_1n_2}^{\beta}(i;i\omega_n)\delta_{n_1m_2} \bigr]<\rho_{n_3n_4}(l)>
$$
\begin{equation}     
+ \bigl[<\rho_{n_1m_2}(i)>\delta_{n_2m_1}
          -<\rho_{m_1n_2}(i)>\delta_{n_1m_2} \bigr] 
          \chi_{n_3n_4}^{\beta}(l;i\omega_n)  \biggr\rbrace.
\label{eq_motion_2}
\end{equation}
}  
Note that the matrix elements $<m|r^{\beta}|m_1>$ are the same
as those appearing in Eq. \ref{ucm}, evaluated for $\vec{R}_i=0$.

Eq. \ref{eq_motion_2} may be solved if one knows the groundstate
densities $<\rho_{m_1m_2}(i)>$.  It is here that our assumption
that all the electrons reside fully in the lowest angular momentum
orbitals for their sites in the groundstate becomes useful:
the form of the densities is simply 
$<\rho_{m_1m_2}(i)>=\delta_{m_10} \delta_{m_20}.$
Furthermore, as shown in Appendix A, it can be proven
that $\sum_lU^{\prime~il}_{n_1n_200} \propto \delta_{n_1n_2}$,
so that Eq. \ref{eq_motion_2} is considerably simplified,
taking the form
\vbox{
$$
i\omega_n \chi_{m_1m_2}^{\beta}(i;i\omega_n) =
\quad\quad\quad\quad
$$
$$
 <0|r^{\beta}|m_1>\delta_{m_20} - 
         <m_2|r^{\beta}|0>\delta_{m_10}  
$$
$$         
+ \bigl[ \tilde{\varepsilon}_i(m_1) - \tilde{\varepsilon}_i(m_2) \bigr] 
   \chi_{m_1m_2}^{\beta}(i;i\omega_n)
$$
$$
+\sum_{n_3n_4} \sum_l \bigl[U^{\prime il}_{0m_1n_3n_4}\delta_{m_2,0}
- U^{\prime il}_{m_20n_3n_4}\delta_{m_1,0} \bigr]
$$
\begin{equation}
\times \chi^{\beta}_{n_3n_4}(l;i\omega_n),
\label{computed_eq_motion}
\end{equation}
}  
where $\tilde{\varepsilon}_i(m)=\varepsilon_i(m) 
+\sum_l U^{\prime il}_{mm00}$.
It is also easy to show that $\chi^{\beta}_{m_1m_2}=0$ if
either $m_1$ and  $m_2$ are both zero, or if both are non-zero.
This is the equation we actually work with, and it is solved
in a way closely analogous to the method developed in
Ref. \onlinecite{cote91}.  We regard $\chi^{\beta}_{m_1m_2}(l)$
as a vector whose components are labeled by $m_1,m_2,l$.
Eq. \ref{computed_eq_motion} may be written schematically
in the form 
$$
\sum_{n_1,n_2,l} \biggl\lbrace
\bigl[ i\omega_n + \Delta\varepsilon_i(m_1,m_2) \bigr]
\delta_{(i,m_1,m_2),(l,n_1,n_2)}
$$
$$ \quad\quad\quad
+ U^D_{(i m_1 m_2),(l n_1 n_2)} \biggr\rbrace 
\chi^{\beta}_{n_1n_2}(l;i\omega_n)
$$
$$
\equiv \bigl[ i\omega_n \delta_{(i,m_1,m_2),(l,n_1,n_2)}
-M_{(i m_1 m_2),(l n_1 n_2)} \bigr] \chi^{\beta}_{n_1n_2}(l;i\omega_n)
$$
$$
= \chi^{\beta~0}_{m_1m_2}(i;i\omega_n)
$$
where $\Delta\varepsilon_i(m_1,m_2) \equiv \tilde{\varepsilon}_i(m_1)
-\tilde{\varepsilon}_i(m_2)$ is the energy required to excite an
electron on site $i$ from the $m_1$ orbital into the $m_2$
orbital when all the other electrons are static, 
$$
U^D_{(i m_1 m_2),(l n_1 n_2)} =
U^{\prime il}_{0m_1n_3n_4}\delta_{m_2,0}
- U^{\prime il}_{m_20n_3n_4}\delta_{m_1,0}
$$
is the change in this energy difference due to the fact that
all the electrons are in fact dynamic, and 
$\chi^{\beta~0}_{m_1m_2}(i;i\omega_n) = <0|r^{\beta}|m_1>\delta_{m_20} - 
<m_2|r^{\beta}|0>\delta_{m_10}$ is the response function in the
absence of both electron-electron interactions and the disorder
potential.  The matrix ${M}$ may be diagonalized numerically;
it will have real eigenvalues $\omega_j$ (which may be shown to come
in pairs of equal magnitudes but opposite signs) and eigenvectors
$V^{j}(m_1m_2l)$.  It can be shown that when regarded as a matrix
in the indices $j$ and $(m_1m_2l)$, $V$ has an inverse $V^{-1}$.
Denoting ${\omega_j}$ as the
diagonal matrix of eigenvalues of $M$, 
the solution to Eq. \ref{computed_eq_motion}
may be written schematically as
$$
\vec{\chi}^{\beta}(i\omega_n) = \vec{\chi}^{0~\beta}(i\omega_n)
V [i\omega_n{\bf 1} - { \omega_j} ]^{-1} V^{-1}.
$$
From this form, it may be seen that the poles of $\chi$, and
hence the energies of collective modes of the system, are given
by the set of eigenvalues $\omega_j$.

As a first example, as well as a check on whether this method
works, we consider the situation of a WC in an undisordered 
environment.  One only needs then to set $\varepsilon_i(m)=0$
for all $i,~m$ in Eq. \ref{computed_eq_motion}, and proceed
as described above.  A histogram of the collective mode energies,
which represents the density of states for the system,
is shown in Fig. 3.  In this calculation, the number of
electrons was $N=225$, the filling factor was set to $\nu=0.2$,
and 5 states per site were retained.  As may be seen, a
broad peak near the origin is accompanied by three larger, well-separated
peaks at higher energy.  The number of such sharp peaks appearing
in the density of states we find in general to be equal to 
the number of orbitals per site above the groundstate 
retained in the calculation. 

The meaning of the sharp peaks may be interpreted if one notes
that their positions are very close to the values of
$\Delta {\varepsilon}_i(m,0)$ used in the equations
of motion.  This means that the collective modes
are only very weakly affected by $U_D$, so that
on may understand the modes as local excitations
in which an electron is excited to a high angular
momentum.  Thus, these peaks may properly be
understood as an analog of Wannier excitons
in tight-binding models of electron systems.
Interestingly, these peaks have been previously
noted in an approach that does {\it not} 
use a harmonic approximation\cite{cote91}.
However, in that approach, the nature of
these excitations was unclear; 
the introduction of the harmonic approximation
allows us to understand why there is structure
in the density of states at these energies.
A careful comparison of the peak
positions for $\nu=0.25$ found in Ref. \onlinecite{cote91}
(see Figs. 2 and 3 of that work) with results
from the method described here shows that the
energies of these peaks are nearly identical
for the lowest energy peaks.

Fig. 4 illustrates the density of states for a
calculation with $N=529$, $\nu=0.2$, and just
two states per sites retained.  As may be seen,
the density of states is precisely what one
expects for a phonon density of states in
a magnetic field: the leading edge of the
phonon density of states rises sharply, consistent
with the $D(\omega) \propto \omega^{1/3}$ expected
for the low-frequency behavior of the density of
states that results from the $\omega(k) \propto k^{3/2}$
dispersion of the WC in a magnetic field\cite{fukuyama75}.
The peak contains a strong cusp structure consistent
with a van Hove singularity, and the width of the peak
is found to decrease with increasing magnetic field,
consistent with the $1/B$ dropoff expected for the
phonon bandwidth\cite{fukuyama75}.

The quantitative agreement of these results with previous
calculations give a strong indication that the QHA works,
and we learn from the above calculations that all the
low-energy modes (i.e., phonons) of the WC can be captured
by retaining just two states per site.  This considerable
simplification allows one to treat fairly large numbers
of electrons for the pinned WC; our largest calculations
contain $N=1024$ electrons.  From this point onward
we will adopt the two-state approximation, and denote
the energy difference between the two states on a site
as $\Delta{\varepsilon}_i(m=1,0)$ as $\Delta{\varepsilon}_i$

Having established that the QHA gives sensible results for
cases where the correct answer is known, we can
now use it to investigate
the case of the pinned WC.   
We defer detailed discussions of the results to Section \ref{sec:results};
however, they can be summarized extremely briefly.  Essentially
all of our center of mass response functions are dominated
by a single collective mode, the lowest excited state arising
for any given disorder realization (e.g. Fig. 1).  
For the weak disorder models
[i.e., $\Delta\varepsilon_i  \ll \omega_B$, where
$\omega_B$ is the bandwidth of the phonon density of states
(Fig. 4)],
the position
of the pinning frequency turns out to be extremely close
to $v_0$, the average pinning
energy per site.  We find
this form of the center of mass response 
for all our disorder models and system sizes studied,
and there is no noticeably trend for the strength of this
one mode to decrease with increasing system size  within the
sizes one may study using this model.  One is led to conclude
that the electromagnetic response of the WC in a strong magnetic
field at zero temperature is extremely sharp, consistent with 
experiment but not with previous theoretical
expectations.  As explained in the introduction, this turns
out to be a consequence of the long-range nature of the
Coulomb interaction.  To see precisely how this result arises, as well
as to assess whether it will survive in the thermodynamic
limit, we need to develop an approach from which one may
learn what happens to this mode for very large systems.
This is the subject of the next section.

\section{Pseudospin Representation}
\label{sec:spin_rep}

The results of the QHA indicate that generically there is a sharp resonance
in the low frequency response of the magnetically induced WC
for the finite size systems one is able to handle with that
method.  While there is no indication within those calculations
that the resonance either broadens or weakens with increasing
system size, one cannot rule out based upon them the possibility
that a broadening does develop at very large system sizes.
Thus one would like to develop models that are analytically
tractable wherein the thermodynamic limit may be taken.

\subsection{Quantum Spin Model}

In this subsection, we will develop a mapping of the electron
system onto an effective spin lattice system, and obtain
couplings between these spins using an expansion in
$l_0/a_0$, with $a_0$ the nearest neighbor distance.
One important advantage of this formulation is
that the resulting couplings are far more tractable
than the matrix elements found in the QHA, allowing
us to make considerable progress analytically.  The spin
waves of the model are the analog of the phonons for the
WC, and it will be shown that the resulting dispersion
of these spin waves is identical to the classical
phonon spectrum of the WC, demonstrating that the
mapping produces sensible results.

We begin with the observation from the QHA analysis that 
essentially all the low energy excitations of the magnetically-induced
WC may be obtained in a model that retains only two states
per lattice site.  This motivates an approach in which one
may wish to consider the on-site degrees of freedom of
each electron as an effective pseudospin, with the $m=0$
angular momentum state representing a ``spin up'' state, and
the $m=1$ a ``spin down'' state. 
(Note that we will assume the real spins of the electrons
are polarized by the magnetic field and do not represent
a low-energy degree of freedom for the system.)
The system is then mapped
onto a quantum magnet, whose interactions we will see are
of a magnetic dipole form, and whose spin waves will
have the same dispersion relation as the phonon
spectrum of the underlying classical electron degrees
of freedom.  Formally, the mapping is given by:
\begin{eqnarray}
a_{0j}^{\dag}a_{0j}-&a_{1j}^{\dag}a_{1j}& \rightarrow 2S^z_j
\nonumber\\
a_{0j}^{\dag}a_{1j} &\rightarrow& S^+_j
\nonumber\\
a_{1j}^{\dag}a_{0j} &\rightarrow& S^-_j.
\label{spin_mapping}
\end{eqnarray}
To write an effective Hamiltonian for these pseudospin degrees of
freedom, it is convenient to employ a multipole expansion of the
electron density, which becomes quantitatively accurate in the
limit $l_0/a_0 \ll 1$ (i.e., $\nu \ll 1$), where $a_0$ is the nearest neighbor
distance.  Neglecting exchange effects, 
the interaction energy is formally
\begin{equation}
U={{e^2} \over 2} \sum_{i \ne j} \int d^2r_1d^2r_2
{{\rho(\vec{r}_1+\vec{R}_i) \rho(\vec{r}_2+\vec{R}_j)}
\over
{|\vec{r}_1-\vec{r}_2|} }.
\label{interaction}
\end{equation}
Writing $\rho(\vec{r}+\vec{R}_i) \equiv \rho_i(\vec{r})$,
the densities may be written in terms of the pseudospin
operators,
\begin{eqnarray}
\rho_i(\vec{r})=|\phi_0^2(\vec{r})|^2[{1 \over 2} + S_i^z]
+|\phi_1^2(\vec{r})|^2[{1 \over 2} - S_i^z]
\nonumber \\
+\phi_0^*(\vec{r})\phi_1(\vec{r})S_i^-
+\phi_1^*(\vec{r})\phi_0(\vec{r})S_i^+,
\label{density_expansion}
\end{eqnarray}
where $\phi_{0(1)}$ are the $m=0(1)$ orbitals localized around
the site of interest (Eq. \ref{orbitals}), and we have
suppressed the explicit site labels $i(j)$.  Substituting
Eq. \ref{density_expansion} into Eq. \ref{interaction}, 
expanding the integrand to second order in $r_{1(2)}/R_{ij}$
where $R_{ij} \equiv |\vec{R}_i-\vec{R}_j|$, and performing the
integrations, one arrives at the formula 
$U = {1 \over 2}\sum_{i \ne j} U_{ij}$, with
\vbox{
$$
U_{ij} = {{e^2} \over {R_{ij}} } + {{3 e^2 l_0^2} \over {2R_{ij}^3} }
-{{e^2 l_0^2} \over {2R_{ij}^3} } \bigl[S_i^z+S_j^z \bigr]
$$
$$
+{{2e^2 l_0^2} \over {R_{ij}^3} } 
\bigl[S_i^{\parallel} \cdot S_j^{\parallel} 
-3(S_i^{\parallel} \cdot \vec{R}_{ij})
(S_j^{\parallel} \cdot \vec{R}_{ij}) / R_{ij}^3 \bigr],
$$
}
where $S_{i(j)}^{\parallel} \equiv (S^x_{i(j)},S^y_{i(j)},0).$
Note that the effective interaction between spins in $U_{ij}$
is of an $XY$ dipole form, which is not surprising
given the fact that a small $r$ expansion for the electron-electron
interaction leads to an electric dipole interaction.
The third term in $U_{ij}$ takes the form of an effective
magnetic field that tends to orient the spins in the
$+\hat{z}$ direction; this reflects the fact that the
$\phi_0$ state is the groundstate of an electron in
a given site if all the other electrons are fixed
in their $\phi_0$ states.  The Hamiltonian for the
effective spin system thus may be written in the form
\begin{equation}
H = \sum_i h_i S_i^z + \sum_{\alpha,\beta} \sum_{i \ne j}
J^{\alpha,\beta}_{ij} S_i^{\alpha}S_j^{\beta}
\label{spin_hamiltonian}
\end{equation}
where $\alpha,\beta=x,y$, $J^{\alpha,\beta}_{ij}=
{{e^2 l_0^2} \over {R_{ij}^3}}[\delta_{\alpha,\beta}
-3R_{ij}^{\alpha}R_{ij}^{\beta}/R_{ij}^2]$,
and $h_i=-\sum_{j (\ne i)}{{e^2 l_0^2} \over {2 R_{ij}^3} } + \varepsilon_i(m=0) - 
\varepsilon_i(m=1)$ is an effective magnetic field,
which may be non-uniform due to the pinning potential.
Note that we have dropped an irrelevant constant
from the Hamiltonian.

The low-energy collective modes of this model are
spin waves, and in the absence of a pinning potential
their dispersion relation should be identical to that
of the phonon modes of the underlying WC from which
the pseudospin model was derived.  We thus begin our
analysis by considering the case $\varepsilon_i(m)=0$.
To derive the spin-wave spectrum, we rewrite the spin
operators in terms of bosonic degrees of freedom\cite{com4}
with the approximate mapping:
\begin{eqnarray}
S_i^z \rightarrow &{1 \over 2}& - b_i^{\dag}b_i \nonumber \\
S_i^+ &\rightarrow& b_i \nonumber \\
S_i^- &\rightarrow& b_i^{\dag} .       \nonumber \\
\label{spin->boson}
\end{eqnarray}

This mapping is an approximate form of the Holstein-Primakoff
transformation\cite{holstein}, expanded for the situation 
\hbox{$<b_i^{\dag}b_i> \ll 1/2$},
where $<\cdot\cdot\cdot>$ is an expectation value for any
of the low energy states that are of interest in the zero
temperature response function.
A simple way to see that $<b_i^{\dag}b_i>$ is small is to write
it directly in terms of the underlying electron creation and
annihilation operators: $b_i^{\dag}=a_{1i}^{\dag}a_{0i}$,
$b_i=a_{0i}^{\dag}a_{1i}$.  One then finds 
$[b_i,b_i^{\dag}]=a_{0i}^{\dag}a_{0i}-a_{1i}^{\dag}a_{1i}$.
If the mapping were exact, this commutator would be unity.
However, to the extent that the groundstate is well-approximated
by electrons in Gaussian orbitals, $<a_{0i}^{\dag}a_{0i}>=1$
and $<a_{1i}^{\dag}a_{1i}>=0$ in the groundstate.  Furthermore,
the lowest excited states -- the spin waves in the 
bosonic language --
are described in terms of the electronic degrees
of freedom by a single particle excited out of the $m=0$
state into the $m=1$ state, averaged with an appropriate phase
factor over all the sites. Thus, the expectation value of
$[b_i,b_i^{\dag}]$ for the low-lying states is the same as that
of the groundstate, up to corrections that vanish in the
thermodynamic limit.  It should be noted that when quantum
(or, at finite temperature, thermal)
fluctuations are important, then one must
work with the exact mapping between spin and boson 
operators\cite{holstein},
and higher order terms in $b_i^{\dag}b_i$ should be retained,
introducing spin wave interactions.  Such terms are the analog
of anharmonic terms in the Hamiltonian for the underlying 
WC degrees of freedom, and we expect them to be very small
at zero temperature for large magnetic fields (small $\nu$),
where placing the electron in Gaussian orbitals at specified
locations that minimize the potential energy
is thought to be an excellent approximation for
the groundstate\cite{maki83}.  From this point onward
we will ignore spin-wave interactions, and our goal
will be to understand why disorder
does not broaden the electromagnetic response
associated with the quadratic Hamiltonian below.

In terms of the bosonic operators, the Hamiltonian 
we thus will consider takes the form
$$
H=\sum_iU_ib_i^{\dag}b_i \quad\quad\quad\quad
$$
$$
+\sum_{i \ne j} {{e^2 l_0^2} \over {R_{ij}^3} }
\biggl\lbrace
-{1 \over 4} (b_i^{\dag} b_j + b_i b_j^{\dag} )
$$
\begin{equation}
-{3 \over 4} \bigl[ (n^{*}_{ij})^2 b_i^{\dag}b_j^{\dag}
                  + (n_{ij})^2 b_ib_j \bigr] \biggr\rbrace.
\label{boson_hamiltonian}
\end{equation}
Here, $U_i=- h_i$, and
$n_{ij}=(R_{ij}^x-iR_{ij}^y)/R_{ij}$ is a complex representation
of the vector direction separating sites $R_i$ and $R_j$.
In the absence of pinning $[\varepsilon_i(m)=0]$, $H$
is diagonalized in two steps.  First, a canonical transformation
from real to momentum space separates out the independent modes:
$$
H=\sum_{\vec{k}} F(\vec{k}) b_{\vec{k}}^{\dag} b_{\vec{k}}
$$
$$
-\sum_{\vec{k}} \bigl[ G(\vec{k}) b_{\vec{k}} b_{-\vec{k}} +
G^*(\vec{k}) b^{\dag}_{-\vec{k}} b^{\dag}_{\vec{k}} \bigr],
$$
where $b_{\vec{k}} = {1 \over {N^{1/2}} }\sum_i e^{i\vec{k} \cdot \vec{R}_i}
b_i$, $F(\vec{k})={1 \over 2} \sum_j {{e^2 l_0^2} \over {R_j^3}}
(1-e^{-i \vec{k} \cdot \vec{R}_j })$, and
$G(\vec{k})={3 \over 4} \sum_{\vec{R}_j \ne 0} {{e^2 l_0^2} \over
{R_j^3} } (n_{\vec{R}_j})^2 e^{i\vec{k} \cdot \vec{R}_j}$.
The diagonalization of the Hamiltonian is then completed with
a Bogoliubov transformation of the form $\gamma_{\vec{k}}=
u_{\vec{k}}b_{\vec{k}}+v_{\vec{k}}b_{\vec{-k}}^{\dag}$,
with $|u_{\vec{k}}|^2-|v_{\vec{k}}|^2=1$ guaranteeing that
$[\gamma_{\vec{k}},\gamma_{\vec{k}}^{\dag}]=1$.  The boson
Hamiltonian then may be written as $H=\sum_{\vec{k}}E_{\vec{k}}
\gamma_{\vec{k}}^{\dag}\gamma_{\vec{k}}$ if one chooses
\begin{eqnarray}
u_{\vec{k}}=\cosh \theta_{\vec{k}} e^{i\phi_{\vec{k}}/2} \nonumber \\
v_{\vec{k}}=\sinh \theta_{\vec{k}} e^{-i\phi_{\vec{k}}/2} \nonumber \\
E_{\vec{k}}=\sqrt{F(\vec{k})^2-4|G(\vec{k})|^2}
\label{diagonalize}
\end{eqnarray}
with $G(\vec{k})=e^{i\phi_{\vec{k}}} |G(\vec{k})|$, and
$\tanh 2\theta_{\vec{k}} = -2|G(\vec{k})|/F(\vec{k})$.

$E_{\vec{k}} $ thus represents the spin-wave dispersion
for this model, and we need to confirm that it has
the same form as 
the underlying electron degrees of freedom.  For the
WC in two dimensions in the absence of a magnetic
field, the phonon dispersion is found by solving the
eigenvalue equation\cite{bonsall77}
\begin{equation}
\sum_{\beta} C_{\alpha\beta}(\vec{k}) e_{\beta} 
= \omega^2(\vec{k}) e_{\beta},
\label{b=0}
\end{equation}
where 
\begin{equation}
C_{\alpha\beta}(\vec{k})=-{{e^2} \over {m^*}}
\biggl\lbrace
\sum_j \bigl( {{3R_j^{\alpha}R_j^{\beta}} \over {R_j^5}}
- {{\delta_{\alpha,\beta} } \over {R_j^3} } \bigr)
\bigl(e^{-i\vec{k}\cdot \vec{R}_j } - 1 \bigr)
\biggr\rbrace
\label{cab}
\end{equation}
and $m^*$ is the effective mass of the electrons.  For every value
of $\vec{k}$, Eq. \ref{b=0} has two eigenvalues, corresponding
to a longitudinal mode $\omega_l(\vec{k})$ and a transverse mode
$\omega_t(\vec{k})$.  In the presence of a magnetic field, the
equations of motion for the electrons mix these two modes. One then
obtains two different normal modes, one dispersing from the
cyclotron frequency $\omega_c=eB/m^*c$, and the other having
the form in a the strong field limit\cite{fukuyama75}
$\omega(\vec{k})=\omega_l(\vec{k})\omega_t(\vec{k})/\omega_c.$
By solving Eq. \ref{b=0} and using Eq. \ref{cab}, one may
show with some algebra that $\omega_l(\vec{k})\omega_t(\vec{k})/\omega_c
\equiv E_{\vec{k}} $.  Thus the spin waves of our pseusospin model
faithfully reproduce the exact phonon spectrum for the WC in
a strong magnetic field, and we see that our model quantitatively
captures the low energy dynamics of the WC.

\subsection{Uniformly Pinned Wigner Crystal}

In this subsection we will analyze the response function of a
WC in which each site has precisely the same pinning potential.
We will find that the power absorption
from a spatially uniform, time-varying electric field
is sharp as a function of frequency.  
This result is hardly surprising
as there is no disorder in this model.  However, it
will serve as the basis for the perturbative treatment in the
next subsection, and so is useful to analyze in some detail.

To introduce uniform pinning in the model, one only needs to
set $h_i=-\sum_{\vec{R} \ne 0} {{e^2l_0^2} \over {2R^3} }
- v_0$ for all the sites $i$ in Eq. \ref{spin_hamiltonian}.
The representation of the spin wave Hamiltonian in terms
of phonon degrees of freedom, and its diagonalization, are
formally identical to the steps used in the last subsection,
provided one makes the replacement
$$
F(\vec{k}) \rightarrow  F(\vec{k}) + v_0 \equiv F_p(\vec{k}).
$$
The resulting (pseudo)spin wave for this model is thus
\begin{equation}
E_{\vec{k}}^p=\sqrt{F_p(\vec{k})^2-4|G(\vec{k})|^2}.
\label{pinned_spin_waves}
\end{equation}
In the limit $k \rightarrow 0$, $E_{\vec{k}}^p \rightarrow v_0$.
Thus for collective modes in which all the electrons move together,
only the center of mass degree of freedom is relevant:
electron-electron interactions may be ignored, and one obtains the
single electron excitation frequency\cite{kohn61}.  The long wavelength
dispersion of Eq. \ref{pinned_spin_waves} may be obtained
by evaluating $F(\vec{k})$ and $G(\vec{k})$ using an Ewald
sum technique\cite{bonsall77}.  One finds
\begin{eqnarray}
F(\vec{k}) &\approx& 2\pi e^2l_0^2  \rho_0 k + O(k^2) \nonumber \\
G(\vec{k}) &\approx& \pi e^2l_0^2  \rho_0 (k_x + i k_y)^2 / k + O(k^2), \\
\label{fg_smallk}
\end{eqnarray}
so that 
\begin{equation}
E_{\vec{k}}^p \approx v_0 + 2\pi e^2l_0^2 \rho_0 k +O(k^2).
\label{psw_smallk}
\end{equation}
The linear dispersion of the pinned collective mode spectrum is
purely a result of the long-range nature of the Coulomb interaction.
For a short-range interaction, the collective mode disperses instead
as $k^2$ from $v_0$.  The linear dispersion, and hence the long-range
nature of the Coulomb interaction, turns out to play a crucial role
in allowing the sharp response of the uniformly pinned system to
survive the introduction of disorder.  This will be discussed
more carefully in the next subsection.

Because one may exactly diagonalize the Hamiltonian in this model,
it is convenient to compute the power absorption using Fermi's
Golden Rule.  For an electric field of the form 
$\vec{E}(t)={\rm Re}~E_0 (\hat{x}+i\hat{y}) e^{i \omega t}$,
one obtains for zero temperature
\begin{equation}
P(\omega) = N \pi e^2 E_0^2 l_0^2 \sum_{n} E^p_{n}
|<n|\gamma_{\vec{k}=0}^{\dag}  |0>|^2 
\delta (\omega-E^p_{n}),
\label{power_absorbed}
\end{equation}
where $|0>$ is the groundstate of the system
(no spin waves), and $|n>$ represents the set of
single pseudospin wave excitations (which in the present
case may be labeled by $\vec{k}$ rather than $n$). 
For the uniformly pinned system, the matrix element
entering Eq. \ref{power_absorbed} is nonvanishing only for
$E^p_{\vec{k}}=v_0$, so that one obtains a delta function
response at $\omega=v_0$, as expected.  

Eq. \ref{power_absorbed}
is a good starting point for a perturbative treatment of
disorder effects.  In particular, the weight arising
from the matrix element $<n| \gamma_{\vec{k}=0} |0>$
must remain proportional to the system size in the
thermodynamic limit for one particular mode $n$ if
the system is to retain the sharp response observed
in Sec. \ref{sec:qha} for an arbitrarily large system.
In the next subsection, we show that, at least for weak
disorder, this is indeed the case.

\subsection{Weak Disorder: Perturbative Treatment}

In this subsection, we will 
formulate a perturbation theory for Eq. \ref{power_absorbed},
it terms of deviations of the pinning potential from 
uniformity.  The point of this analysis is to understand
how the sharpness of the response might survive the
introduction of disorder.   Towards this end, we will
focus on the height of the $\delta$-function response
found in the last subsection, and develop an expression
to the lowest non-trivial order in perturbation theory
to see how much it is decreased by disorder.  The resulting
expression, when disorder averaged, 
will turn out to have a formal, logarithmic
divergence, which we interpret to mean that there
is no true $\delta$-function response in the thermodynamic
limit.  However, the divergence is in fact cutoff
by the system size, and we will see that even a conservative
estimate of the integral for real system sizes indicates that
it is in fact small, so that the perturbation theory
is valid.  The integral allows us to define a
length $L_c$ that is the characteristic size scale
for electrons moving together in phase in the lowest
collective mode, and we will see that $L_c$ is extremely
large compared to real sample dimensions.  

The quantity we will use for our perturbing parameter
is the deviation of the pinning potential from
perfect uniformity, $\Delta U_i$,
which is formally defined by the equation
$h_i=-\sum_{\vec{R} \ne 0} {{e^2l_0^2} \over {2R^3}}
-v_0-\Delta U_i$.   Note from its definition
that $\sum_i \Delta U_i=0$, a property
we will use below.  The 
power absorption (Eq. \ref{power_absorbed}) 
may be written via standard many-body
manipulations\cite {mahan} in terms of a Green's function, in
the form
\begin{equation}
P(\omega) = -e^2l_0^2E_0^2 \omega Im \bigl\lbrace
\sum_{\mu\nu} \sum_{ij} d_{\mu}^0(i)
d_{\nu}^{0}(j)^*G_{\mu\nu}(ij;\omega+i\delta) \bigr\rbrace.
\label{mb_power_absorbed}
\end{equation}
The Green's function matrix entering above is given 
in imaginary time by
\begin{equation}
G_{\mu,\nu}(ij;\tau)=-<T_{\tau} b_i^{(\mu)}(\tau) b_j^{(\nu) \dag}(0) >
\label{greens_function}
\end{equation}
with $\mu,\nu=1,2$, $b_i^{(1)} = b_i$, and $b_i^{(2)}=b_i^{\dag}.$
The c-numbers $d_{\mu}^0(i)$ entering Eq. \ref{mb_power_absorbed}
are those that
diagonalize the pseudospin-wave Hamiltonian Eq. \ref{boson_hamiltonian}
for the $\vec{k}=0$ mode in the uniformly pinned case
(i.e., $d_{1}^0(i) \equiv v_{\vec{k}=0}=0,~d_{2}^0(i) \equiv u_{\vec{k}=0}=1$,
with $u,~v$ given by Eq. \ref{diagonalize}).  Using the method
described in Section \ref{sec:qha} and Eq. \ref{boson_hamiltonian},
the equation of motion for the Green's function 
in imaginary time is found to be 
\begin{eqnarray} 
-i \omega_n \: \left( \begin{array}{c}
G_{11}(ij;i\omega_n) \\
G_{21}(ij;i\omega_n)  \end{array}  \right)
=
-\delta_{ij} \left( \begin{array}{c}
1 \\
0  \end{array} \right) 
\nonumber
\end{eqnarray}
\begin{eqnarray}
+\Delta U_i
\left[
\begin{array}{cc}
-1&0\\
0&1 \end{array} \right]
\left( \begin{array}{c}
G_{11}(ij;i\omega_n) \\
G_{21}(ij;i\omega_n)  \end{array}  \right)
\nonumber
\end{eqnarray}
\begin{eqnarray}
+\sum_{k}
\left[ \begin{array}{cc}
-F_p(\vec{R}_i-\vec{R}_k) & -2G^*(\vec{R}_i-\vec{R}_k) \\
2G(\vec{R}_i-\vec{R}_k) & F_p(\vec{R}_i-\vec{R}_k)
\end{array}\right]
\left( \begin{array}{c}
G_{11}(kj;i\omega_n) \\
G_{21}(kj;i\omega_n)  \end{array}  \right),
\nonumber
\end{eqnarray}
\begin{equation}
\quad\quad
\label{greens_function_eq_motion}
\end{equation}
where
$$
F_p(\vec{R}) = {{e^2l_0^2} \over 2} 
 \sum_{\vec{R}^{\prime} \ne 0}
{1 \over {R^{\prime 3}}} 
\biggl[ \delta_{\vec{R},0} - \delta_{\vec{R},\vec{R}^{\prime}}
 \biggr](1-\delta_{\vec{R},0})+ v_0 \delta_{\vec{R},0}
$$
\begin{equation}
G(\vec{R}) = {{3e^2l_0^2} \over {4R^3}} 
n_{\vec{R}}^2 (1-\delta_{\vec{R},0}). 
\label{FG}
\end{equation}
Note the the quantities in Eq. \ref{FG} are just the discrete
Fourier transforms of the quantities $F_p(\vec{k})$ and
$G(\vec{k})$ defined in the last subsection.
The matrix elements $G_{12}$ and $G_{22}$ may be
found by using $G_{12}(ij;i\omega_n)=G_{21}^*(ij;-i\omega_n)$
and $G_{22}(ij;i\omega_n)=G_{11}^*(ij;-i\omega_n)$.

Eq. \ref{greens_function_eq_motion} may be solved in a
manner closely analogous to that of Section \ref{sec:qha}:
one needs to solve the eigenvalue equation
\begin{eqnarray}
\sum_{k}
\left[ \begin{array}{cc}
-\Delta U_i \delta_{ik}-F_p(\vec{R}_i-\vec{R}_k) 
& -2G^*(\vec{R}_i-\vec{R}_k) \\
2G(\vec{R}_i-\vec{R}_k) 
& \Delta U_i \delta_{ik} + F_p(\vec{R}_i-\vec{R}_k)
\end{array}\right]
\nonumber
\end{eqnarray}
\begin{eqnarray}
\times
\left( \begin{array}{c}
d_1^{(j)}(k) \\
d_2^{(j)}(k)  \end{array}  \right)
= \omega_j 
\left( \begin{array}{c}
d_1^{(j)}(i) \\
d_2^{(j)}(i)  \end{array}  \right),
\label{eval_eqn}
\end{eqnarray}
and then the solution may be expressed in terms of the eigenvalues
and eigenvectors.  It is not difficult to prove several
properties of the solutions to Eq. \ref{eval_eqn} based
on the symmetry of the matrix\cite{com5}.  In particular, one may easily
show that the eigenvalues come in pairs of equal magnitude
and opposite sign, $\pm \omega_j$.  The eigenvector of the
solution with negative $\omega_j$ may be found from the
solution with positive $\omega_j$ with the transformation
$$d_1^{(-)}(k)=d_2^{(+)*}(k), ~d_2^{(-)}(k)=d_1^{(+)*}(k).$$
Because the matrix being diagonalized is not Hermitian,
these two eigenvectors are not in general orthogonal to
one another.  If one imposes the normalization condition
$\sum_k (|d_2^{(j)}(k)|^2 - |d_1^{(j)}(k)|^2)=1$ for
positive eigenvalues ($\omega_j>0$) and insists
that the eigenvectors associated with each pair of eigenvalues
of the same magnitude obey the above relation, 
the inverse
of the eigenvector matrix may be explicitly constructed.
Writing this as $e_{\mu}^{(j)}(k)$ so that 
$\sum_{\mu,k} e_{\mu}^{(j)}(k) d_{\mu}^{(j^{\prime})}(k) 
= \delta_{j,j^{\prime}}$, we find
$$
e_{1}^{(j)}(k)= -d_1^{(j)*}(k), ~ e_{2}^{(j)}(k) = d_2^{(j)*}(k)
$$
for $\omega_j>0$, and
$$
e_{1}^{(j)}(k)= d_1^{(j)*}(k), ~ e_{2}^{(j)}(k) = -d_2^{(j)*}(k).
$$
for $\omega_j<0$.
With this explicit expression for the eigenvector matrix
inverse, it is possible to write the solution to 
Eq. \ref{greens_function_eq_motion} as
$$
G_{\mu\nu}(ik;i \omega_n) = 
\sum_j { { d_{\mu}^{(j)}(i) e_{\nu}^{(j)}(k) } \over
{i\omega_n + \omega_j} }
$$
for $(\mu\nu)=(11)$ and $(21)$.  For $(\mu\nu)=(22)$ and $(12)$,
the expression is the same as above, except one needs to take the complex
conjugate of the numerator.
Combining this with Eq. \ref{mb_power_absorbed}, and noting that
$d_1^0(i)=0$ so that $\sum_{ik} \sum_{\mu\nu} 
 {d_{\mu}^0(i) e_{\mu}^{(j)}(i)^* d_{\nu}^{(j)}(k)^* d_{\nu}^0(k)}$
is purely real, it follows
$$
P(\omega)=e^2l_0^2E_0^2 \omega \pi \sum_j
[\sum_i d_{2}^0(i)e_2^{(j)}(i)^*]
$$
\begin{equation}
\times
[\sum_k d_{2}^{0}(k)^* d_2^{(j)}(k)^*] \delta(\omega-\omega_j) .
\label{overlap}
\end{equation}
for $\omega > 0$.

For a finite size system, Eq. \ref{overlap} describes absorption
by the system into a discrete set of states, with a weight that
may be interpreted as the square overlap of the mode being
excited with eigenvector of the $\vec{k}=0$ mode for the uniformly
pinned case.  In most situations, one expects as the thermodynamic
limit is approached that the weight associated with each mode
vanishes with increasing system size, while the density of modes
increases, to generate an absorption curve that is a 
continuous function of frequency.  What we have found in
Sec. \ref{sec:qha} seems to imply that the lowest mode in spectrum of
the disordered system retains a much larger overlap than all
the other modes, even for large systems.  To understand how this
arises, we need to compute the overlap sum
$[\sum_i d_{2}^0(i)e_2^{(j)}(i)^*]
[\sum_k d_{2}^{0}(k)^* d_2^{(j)}(k)^*]$ in Eq. \ref{overlap}
for the lowest energy mode,
and see how it scales with increasing system size.  Although
an exact calculation of this quantity is not possible for
an arbitrary disorder strength and arbitrarily large system,
we can at least estimate it for weak disorder to see if and when
a finite overlap survives in the thermodynamic limit.

Towards this end, we compute the eigenvectors of the matrix
in Eq. \ref{eval_eqn} to second order in $\Delta U_i$. The method
by which this is done is identical to standard
non-degenerate perturbation theory in quantum mechanics\cite{merzbacher}.
Denoting $(d_1^{(j)}(\vec{R}_1),d_1^{(j)}(\vec{R}_2),...,
d_2^{(j)}(\vec{R}_1),d_2^{(j)}(\vec{R}_2),...) \equiv
{\bf V}^{(j)}$, we may write the result of the calculation
as
\begin{eqnarray}
{\bf V}^{(j)} &=& [1-D(j)]^{1/2}{\bf V}^{(j)}_0 \nonumber \\
&+& \sum_{k~(\ne i)} {\bf V}^{(k)}_0 \biggl[ {{<k|\Delta U |j>} \over
{E_j^{(0)} - E_k^{(0)} }} + O(\Delta U_i^2) \biggr],
\label{evec}
\end{eqnarray}
where ${\bf V}_0^{(j)},~E^0_j$ are the eigenvectors and eigenvalues
in the absence of the perturbation.
The matrix element is
given by
\begin{eqnarray}
<k|\Delta U |j>=
{\bf W}_0^{(k)T} \left(
\begin{array}{cc}
-\Delta U& 0 \nonumber \\
0 & \Delta U \end{array} \right)
{\bf V}_0^{(j)}
\label{def_matrix_el}
\end{eqnarray}
with $\bf{W}_0^{(k)T}$ representing a column vector
of the form
$(e_1^{(j)}(\vec{R}_1),e_1^{(j)}(\vec{R}_2),...,
e_2^{(j)}(\vec{R}_1),e_2^{(j)}(\vec{R}_2),...)$,
and $\Delta U$ in the above matrix is an $N \times N$ diagonal 
matrix with entries $\Delta U_i$ on the diagonal.
The relevant quantity for our purpose is the
``depletion '' $D(j)$, which one finds to be
\begin{equation}
D(j)=\sum_{k~(\ne j)} \biggl|
{{<k|\Delta U |j>} \over
{E_j^{(0)} - E_k^{(0)} }} \biggr|^2
+ O(\Delta U^3).
\label{depletion}
\end{equation}
The zeroth order eigenvectors and eigenvalues are
easily evaluated, as these were already essentially 
found in the last subsection.  Thus the eigenvectors
without disorder are conveniently labeled by a wavevector $\vec{q}$
and a sign $\pm$ denoting whether the positive or negative
eigenvalue for a given wavevector is being referred to.
Thus we use the eigenvalues $E_{j}^0=\pm E_{\vec{q}} =
\pm [F_p(\vec{q})^2-4|G(\vec{q})|^2]^{1/2}$, and
the corresponding eigenvectors have the form
$d_1^{(\vec{q})}(i) = {1 \over {\sqrt{N}}} e^{i \vec{q} \cdot \vec{R}_i}
v_{\vec{q}}$, $d_2^{(\vec{q})}(i) = {1 \over {\sqrt{N}}} 
e^{i \vec{q} \cdot \vec{R}_i} u_{\vec{q}}$, with
\begin{eqnarray}
u_{\vec{q}} &=& u_{\vec{q}}^+ \nonumber \\
&=& \biggl[ {{F_p(\vec{q}) + E_{\vec{q}} } \over
{2 E_{\vec{q}} }} \biggr]^{1/2} e^{i\phi_{\vec{q}}/2} \nonumber \\
v_{\vec{q}} &=& v_{\vec{q}}^+ \nonumber \\
&=&
- {{ 2|G(\vec{q})| } \over {\sqrt{2 E_{\vec{q}} } } }
\biggl[ {1 \over {F_p(\vec{q}) + E_{\vec{q}} } } \biggr]^{1/2}
e^{-i\phi_{\vec{q}}/2} 
\label{uvpos}
\end{eqnarray}
for $E_j^0>0$, and
\begin{eqnarray}
u_{\vec{q}} &=& u_{\vec{q}}^- \nonumber \\
&=&
- {{ 2|G(\vec{q})| } \over {\sqrt{2 E_{\vec{q}} } } }
\biggl[ {1 \over {F_p(\vec{q}) + E_{\vec{q}} } } \biggr]^{1/2}
e^{i\phi_{\vec{q}}/2} \nonumber \\
v_{\vec{q}} &=& v_{\vec{q}}^- \nonumber \\
&=&
\biggl[ {{F_p(\vec{q}) + E_{\vec{q}} } \over
{2 E_{\vec{q}} }} \biggr]^{1/2}
e^{-i\phi_{\vec{q}}/2} 
\label{uvneg}
\end{eqnarray}
for $E_j^0<0$.
Since the factors $d_2^0$ appearing in the power
absorption (Eq. \ref{overlap}) are just the non-zero part
of the eigenvectors above for $q=0$, it is easy to 
verify that the weight associated with excitations into
the lowest collective mode in power absorption is just
$1-D(0)$ (i..e, there will be a contribution to the sum
over modes appearing in Eq. \ref{overlap} proportional
to $1-D(0)$.)  Provided $D(0) \ll 1$, this will
remain a $\delta$-function contribution at zero
temperature, even as absorption into the other modes may merge
into a broad background.  Thus, $D(0)$ measures the depletion
of the sharp pinning mode response found for the uniformly
pinned model.  It should be noted that because of our
choice of system around which we are doing perturbation
theory, there are no corrections to the pinning mode
frequency at $O(\Delta U)$; the first non-vanishing 
correction is of $O(\Delta U^2)$, so that the frequency
of the pinning mode remains close to its uniformly
pinned value.

Using the forms for the uniformly pinned model in
Eq. \ref{depletion}, one finds
\begin{equation}
D(0)={1 \over N} \sum_{\vec{q} \ne 0}
\biggl\lbrace
{{|u_{\vec{q}}^+|^2 |\Delta U(\vec{q})|^2 } \over
{[v_0-E_{\vec{q}}]^2 }}
+ 
{{|u_{\vec{q}}^-|^2 |\Delta U(\vec{q})|^2 } \over
{[v_0+E_{\vec{q}}]^2 }},
\biggr\rbrace
\label{depl_evaluated}
\end{equation}
with $\Delta U(\vec{q}) =  {1 \over {\sqrt{N}}} \sum_i
\Delta U_i e^{i\vec{q} \cdot \vec{R}_i}$.  The second
term in Eq. \ref{depl_evaluated} is always finite and
is small provided $|\Delta U(\vec{q}) / E_q |^2 $
is small.  The first may potentially diverge even for
small $|\Delta U(\vec{q})|$ because $E_{\vec{q}} \rightarrow v_0$
as $\vec{q} \rightarrow 0$, so that there is a vanishing
energy denominator.  However, for a given disorder realization,
by our choice of the system around which we are performing perturbation
theory, we have $|\Delta U(\vec{q})| \propto q^2$ for small
$q$.  The energy denominator, using Eq. \ref{psw_smallk},
behaves as $[v_0-E_{\vec{q}}]^2 \approx (2\pi e^2l_0^2\rho_0 q)^2$,
so that the integral remains finite, and can be small if
$\Delta U(\vec{q})$ is small enough.

This is the central result of this section, and several
comments are in order.  Firstly, the result of Eq. \ref{depl_evaluated}
is only finite in this analysis because of the long-range
nature of the interaction.  For short-range interactions,
$[v_0-E_{\vec{q}}]^2 \propto q^4$, and one ends up with a divergence
no matter how small $|\Delta U(\vec{q})|$ might be.  Such a
divergence indicates that one cannot stop at second order
in the perturbation as we have done here, and that some
self-consistent treatment is called for\cite{fukuyama78,fukuyama78II}.
Under these circumstances, one expects the resulting
response to be broad, as is the case for most pinned
CDW's\cite{gruner}.  As mentioned in the
Introduction, this is borne out by the QHA of
Section \ref{sec:qha}, which shows that a broad
response is indeed obtained if one uses a
screened rather than long-range Coulomb
potential (cf. Fig. 2).
Secondly, in many calculations where one averages over
disorder configurations, the simplest choice of disorder
models (white noise) introduces fluctuations at all length scales,
so that the disorder average 
$\overline{|\Delta U(\vec{q})|^2} \equiv 
\overline{|\Delta U|^2}$ is
independent of wavevector.  Thus, a disorder average
of Eq. \ref{depl_evaluated} would eliminate the zero in
the numerator as $\vec{q} \rightarrow 0$, and introduce
a logarithmically divergent depletion in the thermodynamic
limit, signaling a broadened rather than sharp
response.  

In practice, however, we find that for physically relevant systems
that the response remains sharp.  The reason for this is
that the divergence only arises for truly infinite systems.
We can define a length scale $L_c$ above which the depletion
$D(0)$ is of order 1, so that the $\delta$ function response
becomes significantly broadened.  To do this, one must solve
the equation
\begin{equation}
1={1 \over {\rho_0}} \int_{q > 2\pi/L_c} {{d^2q} \over {(2\pi)^2}}
\overline{|\Delta U|^2}\biggl\lbrace
{{|u_{\vec{q}}^+|^2  } \over
{[v_0-E_{\vec{q}}]^2 }}
+ 
{{|u_{\vec{q}}^-|^2  } \over
{[v_0+E_{\vec{q}}]^2 }}
\biggr\rbrace.
\label{int_eqn}
\end{equation}
We can break up this integral into singular and non-singular
parts as $L_c \rightarrow \infty$, and so write Eq. \ref{int_eqn}
in the form
$$
1 = {1 \over {2\pi\rho_0}} \int_{2\pi/L_c}^{2\pi {\ell} }
{{dq} \over q} {{\overline{|\Delta U|^2}} \over {2\pi e^2 l_0^2 \rho_0^2} }
+ \eta({\ell}),
$$
where we have used Eq. \ref{psw_smallk}.  In the above equation,
${\ell}$ represents a length scale above which the collective mode
dispersion for the uniformly pinned system is accurately
represented by Eq. \ref{psw_smallk} (${\ell} \approx 10a_0$
would probably be sufficiently large), and $\eta({\ell})$
represents the non-singular contribution to Eq. \ref{int_eqn}.
One may now solve for $L_c$, with the result
\begin{eqnarray}
L_c&=&{\ell} \exp \bigl\lbrace
{{(2\pi\rho_0)^3(el_0)^4} \over {\overline{|\Delta U|^2}} }
[1-\eta({\ell})] \bigr\rbrace \nonumber \\
&=& 
{\ell} \exp \bigl\lbrace
{{\nu^3} \over {|\Delta \tilde{U}|^2} }
[1-\eta({\ell})] \bigr\rbrace .
\label{L_c}
\end{eqnarray}
In Eq. \ref{L_c}, $|\Delta \tilde{U}|^2$ is the disorder potential
strength $\overline{|\Delta U^2|}$ 
written in units of $e^2/\kappa l_0$.  For weak disorder,
$\eta({\ell}) \ll 1$.  We will discuss the
interpretation and consequences of
Eq. \ref{L_c} further in Section \ref{sec:results} below;
for now, however, we point out that an estimate of $L_c$
for physically relevant parameters shows that it is extremely
large, much larger than the physical dimensions any real
sample.  This means that in practice, the depletion $D(0)$
will be small for weak disorder, since the sample size
cuts off the divergence at length scales much smaller
than the one at which broadening becomes significant.

It is interesting to consider what would happen if one
were to consider a model in which the displacement of
electrons from their lattice sites was included as an
effect of the disorder.  The essential change is that
the electron centers $\vec{R}_i$ appearing in 
Eq. \ref{greens_function_eq_motion} are no longer on
lattice sites.  For weak disorder, if one neglects
lattice defects such as dislocations and disclinations,
the effect of the lattice deformation may be described
by writing $F_p(\vec{R}_i-\vec{R}_j) 
 \rightarrow F_p(\vec{R}_i,\vec{R}_j) = F_p^0(\vec{R}_i-\vec{R}_j)
 +\delta F_p(\vec{R}_i,\vec{R}_j)$
and
$G(\vec{R}_i-\vec{R}_j) 
 \rightarrow G(\vec{R}_i,\vec{R}_j) = G^0(\vec{R}_i-\vec{R}_j)
 +\delta G(\vec{R}_i,\vec{R}_j)$
in Eq. \ref{greens_function_eq_motion}, where $F_p^0$ and
$G^0$ are the couplings between electron lattice sites
in some perfect reference lattice.  One then may
treat $\delta F_p$ and $\delta G$ as perturbations in
precisely the same manner as we treated $\Delta U_i$
above.  The resulting depletion $D(j)$ has precisely
the same form as in Eq. \ref{depletion},
with the only difference being that the matrix appearing
in the definition of
$\langle k | \Delta U | j \rangle$ now has off-diagonal
matrix elements due to $\delta F_p$ and $\delta G$.
The resulting expression for the depletion of the lowest
collective mode $D(0)$ has the same form as Eq. 
\ref{depl_evaluated}, although the precise form
of $|\Delta U(q)|^2$ will be more complicated than
for the diagonal disorder model.  The important point,
however, is that the energy denominators in the perturbation
theory will be unaffected, so that one still expects 
only a weak logarithmic divergence if the disorder is
not too strong, and a resulting sharp electromagnetic
absorption for a finite size system.  This demonstrates,
albeit {\it a posteriori}, that our assumption of 
a diagonal disorder model does not alter the qualitative
physics of the system, at least for weak disorder.

As is clear from the
above discussion, the precise results for the length scale $L_c$ depend
upon the 
disorder model one uses for $\Delta U(\vec{q})$.
In the next section, we introduce a simple model
for this 
due to imperfections in the interface upon which the
2DEG resides.

\section{Interface Pinning Model}
\label{sec:disorder}

In both the QHA and the perturbative analysis described above, one
must make a specific choice for the (disorder) pinning potential.
In the Section \ref{sec:results} below we present results for
three different models.  The first is a simple white-noise type
model, in which some fixed fraction of the electrons are given an
excitation energy $v_0+\Delta U_i=U_p$ between the $m=0$ and
$m=1$ orbital states, and all other sites are unpinned.  The simplicity
of this model allows a clear comparison of results obtained
from the QHA and the perturbative analysis.  A second model
we investigate is based on the idea that charged impurities are
likely to be present in the spacer layers of heterojunction systems, 
and that
some of these impurities, if close enough the the 2DEG, will substitute
for electrons in the lattice\cite{ruzin92}.  By choosing a
small number of lattice sites to have extremely large values
of $U_p$, we can model this type of disorder within
the QHA.  As will be seen, this leads to a band of
collective excitations well above the phonon band, corresponding
to localized excitations of the strongly pinned electrons.  At
low excitation frequencies, these electrons are essentially
immobile, and so behave as if they were not degrees of freedom.
This is by definition a strongly pinned system; yet we will 
see in Section \ref{sec:results} that the resulting pinning
frequency is extremely small, and that an unacceptably large
number of such charged impurities must be present in the
system to account for the experimentally observed magnitude
of the pinning frequency.

We thus focus on a pinning mechanism which, to our knowledge,
has not been previously discussed in the context of the 
magnetically induced WC: interface disorder.
The interfaces between GaAs and AlAs at which the 2DEG's reside
are by design of very high quality in samples such as those of
Refs. \onlinecite{li97} and \onlinecite{mellor97}.  
Nevertheless, they cannot be perfect, and it is generally
accepted\cite{interface} that such interfaces can only be defined to within
a single lattice constant of the host semiconductors. 
A simple idealization of this is to model the interface
as a series of pits and/or terraces, with the 
height of the interface fluctuating randomly up and down
as illustrated in Fig. 5.
The typical scale of the pits and terraces formed by the
imperfect interface are generally thought\cite{interface} to be 
of the order of several tens of Angstroms.  For the hole
samples of Refs. \onlinecite{li97} and \onlinecite{mellor97},
the interface has an additional corrugated structure with
a size scale of 32\AA~and a depth of 10.2\AA.\cite{engel97,notzel92}.

There are several reasons for believing that
this form of disorder may be important in the experiments of Refs.  
\onlinecite{li97} and \onlinecite{mellor97}.  Firstly, the 
magnetic field at which one first sees a clear resonance at
the lowest available temperatures is roughly 8T, corresponding
to a magnetic length of $l_0 \sim 90$\AA, a very reasonable
length scale below which one might expect the interface
structure to trap electrons.
If we model the pits in the interface
as flat depressions of depth $\Delta z$ as
in Fig. 5, one can estimate the trap potential for a large
pit by assuming the electric field between the 2DEG and
the charged remote donors is uniform\cite{ando82}.  In this
case the the potential gained from placing an electron in
a pit that is much larger than the
magnetic length is $\Delta V=2\pi\rho_0 e^2\Delta z/\kappa
\approx 4.35K$ in temperature units; $\kappa \approx 12$ here is
the dielectric constant of the GaAs host for the electrons,
and we have taken $\Delta z = 10\AA$.
It should be noted that 
for experimentally accessible magnetic fields, a typical
pit size will generally be much smaller
than the magnetic length, so that the pinning
energy of an electron trapped in a single pit will
be of order $\Delta V s^2/l_0^2$, considerably
smaller than the maximum possible value
of $\Delta V$.

Because of the roughness of the interface, the WC
in general will distort slightly so that some or
all of the electrons may take advantage of the interface
potential.  Ultimately some fraction
of the electrons will find equilibrium centers for their
(groundstate $m=0$) Gaussian orbitals that are particularly
low in energy.  Because the first excited ($m=1$) state
of each orbital is spatially far from the center of the
$m=0$ state (in the sense
that $l_0$ is larger than the average pit size
$s_0$), the former
state will not be correlated with
the disorder potential, and thus is not on average
lowered in energy as is the $m=0$ state.  
It should also be kept in mind that for a given 
groundstate configuration, the energy of 
an excitation of a {\it single}
electron from the $m=0$ state to the $m=1$ state,
keeping all the other electrons fixed, will have
its largest contribution from the electron-electron
interaction rather than from the interface disorder.  
Provided the disorder is not
too strong, the potential well in which an individual electron
resides is thus to a first approximation circularly
symmetric, so that the expansion in
terms of angular momentum states used in this work 
is sensible.

To fully specify the interface disorder model, we need to find
the energy due to the disorder to excite an electron  
out of its groundstate
orbital ($m=0$) into the lowest excited state ($m=1$) for
each site $i$, 
$\varepsilon_i(m=1) - \varepsilon_i(m=0) = v_0 +\Delta U_i$.
The distribution of $\Delta U_i$ depends upon the assumptions
one makes about the surface morphology of the interface, as 
well as whether other pinning sources besides interface roughness
are present in the system.  A fuller accounting of various 
possible models will be discussed elsewhere\cite{unpublished};
here we will focus on one reasonable possibility that 
reproduces both the magnetic field and density
dependence of the resonance frequency observed in 
the experiments of Ref. \onlinecite{li97}.

In this model we make the plausible assumption that the strongest
pinning in the hole samples comes from defects and disorder
in the interface corrugations of these systems.  (It may be shown
that perfectly regular corrugations of this sort lead only
to very weak pinning\cite{unpublished}.)  As a simple model
of this, we assume that the surface has pits of size scale
$s_0 = 30\AA$ and depth $\Delta z = 10\AA$ with a surface density
$n_i$.  We will assume the average distance between pits $1/\sqrt{n_i}$ is
smaller than the nearest neighbor separation of the electrons
$a$ but larger than the magnetic length $l_0$.  This means
that each unit cell of the WC contains several pits in which
an electron might be trapped, but when trapped an electron
wavepacket covers one and only one pit.  In what follows we
will choose the density $n_i$ to fit the experimental data
at the lowest fields for which a resonance is observed; we then
need to go back and check that the resulting pit density is
consistent with $l_0<1/\sqrt{n_i}<a$.

To estimate the density of pinned electrons, we use the collective
pinning approach that has been very successful in the context
of vortices in superconductors\cite{larkin75} and charge density
waves\cite{fukuyama78,normand92}.  The basic idea is to find
a size scale $R_c$ of correlated domains, whose positions
may adjust more or less independently of one another to
take advantage of the pinning potential, at the expense of
lattice distortion energy.  The average number of pins that
may be found under electron wavepackets in a single domain if
placed at a random location is given by $N_{pin}=\rho_0 R_c^2\pi l_0^2 n_i$;
if we vary the position of the domain a distance of order
the disorder correlation length (in this case, $1/\sqrt{n_i}$),
one expects fluctuations in the number of pinned electrons of
order $\sqrt{N_{pin}}$.  Since the energy gained by an electron
wavepacket when sitting over a pit is $\Delta V s_0^2/l_0^2$, 
in analogy with the Fukuyama-Lee-Rice model for charge density waves,
the energy
per electron gained from the disorder potential 
$u^{FLR}_{p}$ is
$$
u^{FLR}_{p} \approx \Delta V {{s_0^2} \over {l_0R_c}}
\bigl[{{\pi n_i} \over {\rho_0}} \bigr]^{1/2}.
$$

The introduction of such distortions inevitably leads to a
lattice distortion of amplitude $1/\sqrt{n_i}$ over a distance
$R_c$.  Following Ref. \onlinecite{normand92}, we assume
the distortions are purely transverse, and so estimate
the distortion energy per electron to be
$$
u^{FLR}_d \approx {{\mu} \over {n_iR_c^2\rho_0}}
$$
where $\mu$ is the shear modulus of the lattice.  Minimizing
the total energy $u^{FLR}_{d} - u^{FLR}_{p}$ with respect
to $R_c$, we arrive at the optimum domain size
\begin{equation}
R_c = {{2\mu l_0} \over {\Delta V s_0^2 (\pi n_i^3\rho_0)^{1/2}}}.
\label{rc}
\end{equation}
The resulting pinning energy per particle is found by substituting
this value of $R_c$ into our expression for $u^{FLR}_{p}$, leading
to the expression
$$
u^{FLR}_{p} \approx {{\pi \Delta V^2 s_0^4 n^2_i} \over {2\mu l_0^2}}.
$$
To use this, we need an explicit form for the shear modulus.  This
can be deduced from the transverse phonon dispersion relation
of the WC in the absence of a magnetic field, which has the
form $\omega_T(q)^2 = {{\mu} \over {m^* \rho_0}} q^2$,
where $m^*$ is the electron effective mass.  
The transverse mode eigenfrequency has been shown\cite{bonsall77} 
in the classical harmonic approximation to
be given approximately by
$$
\omega_T(q)^2 = 0.036 \omega_p^2 (aq)^2
$$ 
with $\omega_p^2=4\pi e^2/\sqrt{3}m^* \kappa a^3$.  Using the above
equation, one arrives at the estimate
\begin{equation}
\mu \approx 0.3{{e^2} \over {\kappa a^3}}.
\label{shear_modulus}
\end{equation}
The resulting pinning energy per particle is finally given by
\begin{equation}
u_p^{FLR} \approx 5.24 {{\kappa a^3 \Delta V^2 s_0^4 n_i^2}
\over {e^2 l_0^2}}.
\label{pinning_energy}
\end{equation}

In the perturbative regime, where the effective depth 
of the individual pinning centers, $\Delta V s_0^2/l_0^2$, is
much smaller than the bandwidth of the phonon density of states,
the pinning frequency is just given by the average binding
energy per site, so that one finds 
$\omega_{pin} \approx v_0 \approx u_p^{FLR}$,
with $u_p^{FLR}$ given by Eq. \ref{pinning_energy}.  Note within
this approximation, $\omega_{pin} \propto l_0^{-2} \propto
B$, which leads to a pinning frequency that {\it increases} with
magnetic field, as seen in experiment.  It is also interesting
to note that $\omega_{pin} \propto a^3 \propto 1/\rho_0^{3/2}$, which
is also consistent with experimental results\cite{liunpub}.
We will obtain quantitative estimates for $v_0$ in the next
section using the above expressions.

When the pinning potential for an electron trapped by a pit
approaches the bandwidth of the phonon density of states,
the perturbative estimate above becomes quantitatively
inaccurate, and one needs to perform a QHA calculation 
to obtain reliable results for $\omega_{pin}$.  To do
this, we need the probability that a given electron
will be trapped in a pit, which in this approach is
given by $u^{FLR}_p/\Delta V ({{s_0^2} \over {l_0^2}})$.  
We are then led to
consider a distribution of pinning energies in
which most electrons are unpinned, and a small fraction
are pinned with an effective potential  
$v_0+\Delta U_i = \Delta V s_0^2/l_0^2$.
The resulting model is then formally identical to the white
noise model mentioned above.

\section{Results}
\label{sec:results}

In this section, we describe in more detail the results of
our study.  As has been emphasized, the electromagnetic responses
computed in the QHA in all the models we have
studied are qualitatively the same: one finds a single
sharp line dominating the absorption spectrum.  This line
is very robust in that its weight shows no discernible
decrease with increasing system size, and for models
in which the pinning potential of the individual electrons
is small compared to the width of the phonon density of
states (Fig. 4), the frequency of the resonance occurs
at the average excitation energy per electron, as expected
from the perturbative analysis of Section \ref{sec:spin_rep}.
We begin by describing in more detail the results for
disorder that is in this sense weak.

\subsection{Pinning by Weak Disorder}
\label{subsec:weak_disorder}

A particularly useful model for comparison 
of the QHA and perturbative analysis
is one in which a fixed fraction $n_{pin}=N_{pin}/N$ of 
randomly chosen electrons
is pinned, with each such site assigned the same pinning potential
$v_0+\Delta U_i  = U_p$.  
According to the perturbative analysis,
the pinning frequency $\omega_{pin} \approx n_{pin}U_p$, regardless
of the precise distribution of pinned sites.  
Figs. 1 and 2 show typical results for this model,
with half the sites pinned, using $U_p = 0.01e^2/\kappa l_0$.
Fig. 6 illustrates
the values of $\omega_{pin}$ as computed in the QHA for
5 different disorder realizations each at several different
values of $v_0+\Delta U_i=U_p$, 
for a smaller density of pinned sites (10\%), and $\Delta U_i =0$
for all other sites,
along with the prediction of the
perturbative analysis (dotted line).  For small $U_p$ the 
agreement of the two approaches is quite good.
Furthermore, there is almost no variation in $\omega_{pin}$
for a fixed value of $U_p$ when it is small as the precise
realization of the pinned sites is changed.  The interpretation
of this observation is that, for small $U_p$, 
$\omega_{pin}$ is determined
almost exclusively by the restoring force on the WC when
only the center of mass coordinate is moved with the
lattice itself undistorted.  Lattice distortions do introduce
some deviation in the resonance frequency, generally pushing
it below the value expected from the lowest order perturbation
theory, in a way that {\it does} depend on the precise
disorder realization as may be seen in the figure.  However,
for any given disorder realization, a single mode always
dominates the response; broadening is never observed for
a fixed disorder configuration.

We note that with higher order corrections to the perturbation
theory of Section \ref{sec:spin_rep}, more accurate predictions
of $\omega_{pin}$ can be made for a given disorder realization.
For example, one of the configurations in Fig. 6 with 
$U_p=0.003$ has $\omega_{pin}=2.32 \times 10^{-4}$.
(The numbers for both $\omega_{pin}$ and $U_p$ here
are in units of $e^2/\kappa l_0$.)  The lowest
order perturbation theory predicts $\omega_{pin} \approx
\omega_{pin}^{(0)} = 3 \times 10^{-4}$ at this density of
pinning sites, which is in error by approximately 35\%.
Using the methods of Section \ref{sec:spin_rep}, the
second order correction may be computed for this
particular disorder realization, with the result
$\omega_{pin} \approx
\omega_{pin}^{(0)} + \omega_{pin}^{(2)} = 2.19 \times 10^{-4}$,
which is within 5.6\% of the QHA result, a considerable
improvement.  (Note that the first order correction
in $\Delta U_i$, $\omega_{pin}^{(1)}$, precisely vanishes
in our perturbative approach.)  Another significant point, as stated
above, is that the weight of this single mode shows
remarkably little size dependence.  For example, using the parameters
relevant to Fig. 1, the power absorption per electron at
the frequency of the sharp peak in the inset of Fig. 1 is
proportional to $\chi_{xx}(\omega_{pin})\omega_{pin}/N=
0.006209$, where $N=1024$ for this calculation.  For 
precisely the same system parameters, but
$N=529$, one finds $\chi_{xx}(\omega_{pin})\omega_{pin}/N=
0.006200$, a slight {\it decrease} for the smaller system size.  This
decrease is almost certainly related to the fact that
the disorder realizations in the two calculations are
inevitably different, rather than to any systematic increase
with increasing system size.  Nevertheless, this
result illustrates that one cannot discern a
decreasing weight in the sharp response with increasing
system size for the values of $N$ one may handle in the
QHA.
 
The important question remains: to what extent, and under
what conditions, does this response remain sharp for systems
of experimentally relevant sizes?  To address this question, we turn
to the perturbative treatment of Section \ref{sec:spin_rep}.
It was shown there that it is convenient to compute the
power absorption by starting from a uniformly pinned state
-- i.e., a pinning potential that is the same for
all electrons in the system -- and computing perturbatively
the corrections to this absorption spectrum due to
the fact that the pinning potential is not truly uniform.
For the uniformly pinned system, it is not at all surprising
that energy absorption from a spatially homogeneous (time-dependent)
electric field is dominated by a single mode: the
collective modes have a well defined wavevector $\vec{k}$,
and only the $\vec{k}=0$ mode can couple to the electric
field.  In principle this $\vec{k}=0$ mode is mixed in among
all the modes when disorder is introduced.  Thus,
as discussed more carefully in Section \ref{sec:spin_rep}, we are
led to ask whether there is a finite overlap between
the  $\vec{k}=0$ collective mode for the uniformly pinned
case and the lowest frequency collective mode in
a disordered system in the thermodynamic limit.

In Section \ref{sec:spin_rep} we showed that this overlap
may be written in the form $1-D$, with $D$ given to second 
order in perturbation theory by 
\begin{equation}
D={1 \over {\rho_0}} \int {{d^2q} \over {2\pi}}
\biggl\lbrace
{{|u_{\vec{q}}^+|^2 |\Delta U(\vec{q})|^2 } \over
{[v_0-E_{\vec{q}}^p]^2 }}
+ 
{{|u_{\vec{q}}^-|^2 |\Delta U(\vec{q})|^2 } \over
{[v_0+E_{\vec{q}}^p]^2 }}
\biggr\rbrace,
\label{depl_evaluated_2}
\end{equation}
with $\Delta U(\vec{q}) =  {1 \over {\sqrt{N}}} \sum_i
\Delta U_i e^{i\vec{q} \cdot \vec{R}_i}$, and 
$\Delta U_i$ is the deviation of site $i$ from the 
site-averaged value of the pinning potential, $v_0$.
Explicit expressions for $u_{\vec{q}}^+$ and 
$u_{\vec{q}}^-$ are given in Eqs. \ref{uvpos} and \ref{uvneg},
and $E_{\vec{q}}^p$ is the dispersion relation for
the uniformly pinned system, Eq. \ref{pinned_spin_waves}.
For small $q$, as shown above (Eq. \ref{psw_smallk})
$$
E_{\vec{q}}^p \approx v_0 + 2\pi e^2l_0^2\rho_0 q +O(q^2),
$$
so that $D$ diverges in the thermodynamic limit, if
$|\Delta U(\vec{q})|^2 \rightarrow \overline{|\Delta U|^2} > 0$
as $q \rightarrow 0$, as is typically the case
for white noise potentials such as the one studied here.
The meaning of this divergence is that essentially
all the weight of the $\vec{q}=0$ mode for the uniformly pinned
system has been depleted
by the disorder from the lowest energy collective mode,
and is distributed
among the other collective modes.  However, in practice
$D$ is only divergent in the
thermodynamic limit, since the integral in  Eq. \ref{depl_evaluated_2}
has an infrared cutoff $q_{min}=2\pi/L$, where $L$ is the
linear dimension of the system size.  The depletion
becomes significant ($D \approx 1$) for system sizes
$L > L_c$, which in Section \ref{sec:spin_rep} was found to
be (Eq. \ref{L_c})
\begin{equation}
L_c
\approx
{\ell} \exp \bigl\lbrace
{{\nu^3} \over {|\Delta \tilde{U}|^2} }
 \bigr\rbrace,
\label{L_c_simple}
\end{equation}
where $|\Delta \tilde{U}|^2$ is the disorder averaged 
square potential evaluated in units of $e^2/\kappa l_0$
for $q \rightarrow 0$, and $2\pi/\ell$ is a wavevector below which
the small $q$ expansion for the uniformly pinned collective
mode spectrum (Eq. \ref{psw_smallk}) becomes reasonably
accurate; presumably $\ell \approx 10a_0$, with $a_0$ the interelectron
lattice spacing.  

For system sizes $L \ll L_c$, the depletion remains small, so that
a single sharp resonance should still be present.  This means that
one may interpret $L_c$ as a length scale for which the electrons
in an infinite sample move together coherently.  
To estimate the width of the resulting resonance,
we note that the fluctuations in the pinning potential averaged
over a length scale of $L_c$ will be $ n_{pin} \bigl[1 \pm 
{{a_0} \over {L_c}} \bigr] \bigl(\overline{\Delta U^2}\bigr)^{1/2}$, 
so that the effective
$Q$ of the resonance would be ${{L_c} \over {2a_0}}$.
If $L_c$ is very large, clearly this leads to an extremely
narrow resonance.  In fact, even a conservative estimate
of $L_c$ shows that it is larger than the physical dimensions
of any real sample.  Taking 
$\bigl(\overline{\Delta U^2}\bigr)^{1/2} \approx 10 \omega_{pin}$, so that
$\Delta \tilde{U}
\approx 10\omega_{pin} \kappa l_0/e^2 \approx 0.014$, where
$\omega_{pin}$ is the observed\cite{li97,mellor97} resonance frequency 
of 1.25GHz, and we used a magnetic length $l_0 \approx 81\AA$
appropriate for a 10T magnetic field.  (We expect that 
$|\Delta \tilde{U}|$ is considerably smaller than this.)
At this field, the
filling factor for the experimental densities is $\nu=0.22$,
so that $L_c \sim 10^{24} a_0$, far larger than any real sample.
The conclusion then is that $L_c \gg L$ in Refs. \onlinecite{li97,mellor97},
and the electrons at zero temperature react essentially as a single
domain in response to the (spatially uniform) electric field.
Furthermore, the depletion $D$ is in fact small in spite of
the formal divergence when $L \rightarrow \infty$ 
at second order in perturbation theory;
this accounts for the sharpness of the measured response
at the lowest temperatures\cite{li97,mellor97}.

Finally, it must be emphasized that this effect arises purely due
to the long-range nature of the Coulomb interaction.  For short
range potentials, $E^p_{\vec{q}} -v_0 \sim q^2$, and the
divergence in Eq. \ref{depl_evaluated_2} is much stronger.
This observation leads to another interpretation of
the result: for short-range interactions, the phonon
density of states for a uniformly pinned system
at small frequencies is much larger
than is the case for long-range interactions.
(In the former case the phonon density of states jumps at
the band edge, whereas in the latter it rises linearly
from zero.)  Thus, if one displaces the center of mass 
of the system (i.e., creates a $q=0$ excitation),
for Coulomb interactions there are very few states into which
the disorder can scatter this excitation.  The calculations
in this work
demonstrate that the suppression of the phonon density of states
at the band edge by the Coulomb interaction is sufficiently
strong to leave a finite oscillator strength of the pure
center of mass (q=0) mode in the lowest frequency collective mode
when disorder is included. 

The observation that Coulomb interactions are crucial to getting
the sharp response
is consistent with the results of the QHA: Fig. 2 illustrates
response for a single disorder realization when the Coulomb
interaction is screened.
For this calculation, we took an electron-electron
interaction of the form
$$
v(\vec{q})=2\pi e^2/\kappa \sqrt{q^2+q_c^2},
$$
with $q_c=2.0l_0^{-1}$, $N=225$, and the parameters are
otherwise the same as for Fig. 1.
The broadening even for this relatively
small system is already apparent.

\subsection{Pinning by Charged Impurities}
\label{subsec:impurities}

A commonly accepted model for understanding the magnitude of the
depinning threshold observed in DC voltage-current 
measurements\cite{goldman90,glattli90,li91} is one in
which charged impurities close to the 2DEG become incorporated
in the lattice as substitutions for electrons\cite{ruzin92}.
This is a paradigm for a strongly pinned system\cite{fukuyama78},
where certain locations of the lattice are essentially ``nailed
down'', and eliminated as degrees of freedom.
In CDW systems, the pinning frequency for this type of disorder
is estimated as the frequency of a phonon mode for the pure system,
evaluated at wavelength $q \sim 2\pi/d$, where $d$ is a typical
separation between pinned sites.  Using the QHA, we can 
quantitatively investigate this model to test whether it
can account for the $\sim$1GHz resonance as well. 

The method for emulating the charged impurity model within the
QHA is to choose a small number of sites for which $\Delta U_i$ is
larger than the width of the phonon density of states.  A typical
collective mode density of states arising from this type of
disorder realization is illustrated in Fig. 7.  Two peaks
emerge, one at high frequency above the resonance frequency
of the individual pinned sites, the other shifted slightly upward from
the phonon density of states for the pure system.  It is easily
checked that the motion of the electrons in the lowest energy
modes leave the pinned electrons stationary, so that the magnitude
of the pinning potential on these sites is irrelevant for these
modes; in practice these electrons are removed as degrees
of freedom.

Fig. 8 illustrates the frequency of the lowest collective mode
as a function of pinned site density, with several different
disorder realizations.  As in all the results of this work,
the electromagnetic absorption is dominated for each realization
by only the lowest mode\cite{com6}.  At low densities, the
frequency is approximately linear in the pinning site density,
a result quite reminiscent of what was found for the weak
pinning model of Section \ref{subsec:weak_disorder} above.
The magnitude of the pinning frequencies found is surprisingly
small; if one extrapolates the linear behavior for small
pinning densities, we find that $\sim 13\%$ of the sites
would have to be pinned in order to achieve the 1.25GHz
resonance seen at $\nu=0.2$.  Such a high density of
charged impurity substitutions near the 2DEG is clearly
inconsistent with the high mobilities these samples exhibit
in zero magnetic field.  Furthermore, the pinning frequency
is a monotonically decreasing function of field in this
model, contrary to experimental observation.  It thus seems
quite clear that this model cannot be the primary pinning
source relevant in electromagnetic absorption.

The behavior of the pinning mode at low impurity
densities is surprising in its similarity to the results
of the weak pinning model above, given that this is
intrinsically a strong pinning mechanism.  Indeed, the
results of a strongly pinned CDW estimate grossly overestimate
the pinning frequency at the densities we have studied.
For example, if we assume a pinned site fraction
$n_{pin} =0.05$, then for filling factor $\nu=0.2$
the strong pinning estimate yields an expected pinning
frequency greater than $7 \times 10^{-3}e^2/\kappa l_0$,
while the frequency found in the QHA is roughly
$5\times10^{-4}e^2/\kappa l_0$.  This great disparity
between the strongly pinned CDW estimate and the 
result of a more realistic calculation we believe
exemplifies the limits of using CDW theory to
quantitatively analyze properties of the magnetically
induced WC.  Indeed, we believe the tendency for
the system at low $n_{pin}$ to behave so much like
a weakly pinned WC may be understood if one
thinks of a strongly pinned electron as a charged impurity to which
a vacancy in the WC has become bound.  One then may
think of the strongly pinned WC as a weakly pinned
WC with point defects, and it is likely that a
perturbative analysis of this system as in Section \ref{sec:spin_rep}
would explain the sharpness of the response found
from the QHA.  If so, the strongly
pinned WC for the purposes of ac response is
equivalent to a weakly pinned,
defective WC.

\subsection{Interface Pinning Model}

We now turn to the results of the interface pinning
model defined in Section \ref{sec:disorder}.  As a first
estimate, we use the perturbative result
$\omega_{pin} \approx v_0 \approx u^{FLR}_p$, with
$u^{FLR}_p$ given by Eq. \ref{pinning_energy}.
As discussed in Section \ref{sec:disorder}, 
we set $s_0 = 30\AA$,
$\Delta V = 4.3K \approx
90GHz$, and take $a=480$\AA, and $\kappa=12$, as is appropriate
for the experiments of Ref.~\onlinecite{li97}.
To estimate the density of pits $n_i$, 
we set $v_0=u^{FLR}_p \approx 1$GHz at $B=8T$ \cite{li97}, for
which $l_0 \approx 90$\AA, and use Eq. \ref{pinning_energy}
to solve for $n_i$; the result is $n_i \approx 2.5 \times 10^{11}cm^{-2}$.
The resulting average distance between pits is $1/\sqrt{n_i}
\approx 200$\AA, which clearly falls into the
range of validity of our model, $l_0 < 1/\sqrt{n_i} < a$.
We note with this estimate of $n_i$, the (Fukuyama-Lee-Rice)
correlation length $R_c$ with the use of Eqs. \ref{rc} 
and \ref{shear_modulus} is
found to be $R_c \approx 7.5a$.  Fig. 9
illustrates the result of the perturbative estimate
(solid line) as a function of magnetic field for this
set of parameters.

The actual potential for sites that are pinned turns out
to be large enough that the perturbative approach 
overestimates the value of $\omega_{pin}$ noticeably.
We thus turn to the QHA to get a more accurate estimate
of the pinning frequency.  As discussed above, this
requires us to find the what fraction of electrons
is pinned; for the above parameters, this works out
to $n_{pin} = N_{pin}/N = 
u_p^{FLR}/\Delta V {{s_0^2} \over {l_0^2}} \approx 10$\%.
It is interesting to note that, since $u_p^{FLR} \propto l_0^{-2}$,
the fraction of pinned electrons is independent of magnetic
field.  Each pinned electron has an excitation energy of
$U_p = \Delta V {{s_0^2} \over {l_0^2}}$ for these 
parameters.  We can then proceed with the QHA 
precisely as in the calculations of Section
\ref{subsec:weak_disorder} above.  
Using the parameters adopted for the perturbative
approach, we find at $B=8T$ in the QHA
$\omega_{pin}=0.84$GHz, slightly below the experimental
value of 1GHz.  This indicates that we should raise our
estimate of $n_i$, since its value was chosen
to match the experimental result for this particular
magnetic field.  We find that the pinning frequency
$\omega_{pin} \approx $1GHz in the QHA if the fraction of pinned
sites $n_{pin}$ is raised to $11\%$;
this can be achieved if we assume a pit density of
$n_i=3.0 \times 10^{11}~cm^{-2}$.  The results of this
calculation are illustrated for several values of magnetic
field in Fig. 9.  
As in Section \ref{subsec:weak_disorder}, the perturbative
result overestimates the pinning frequency by an increasingly
large amount as the value of $U_p$ at the pinned sites
increases; this leads to the sublinear growth of $\omega_{pin}$ with
magnetic field that is apparent in the QHA result.
It is interesting to note that experimentally the
variation of $\omega_{pin}$ with $B$ is indeed found to
be sublinear\cite{li97}.

\section{Discussion}

The primary result of this work, that the zero temperature response
of a pinned, magnetically induced two-dimensional WC to a spatially
uniform, time-dependent electric field is sharp, to our knowledge
is unanticipated in the literature.  As we have seen, the key reason
for this result is the long-range nature of the Coulomb potential.
Early studies of a two-dimensional CDW system in a magnetic 
field\cite{fukuyama78II} did note that for weak disorder potentials
a calculation of the (static) CDW domain size diverges when a
$1/r$ interaction is included.  In Ref. \onlinecite{normand92},
it was argued that this divergence maybe removed be a careful
treatment of the different energy scales for longitudinal and
transverse distortions of the crystal.  This treatment computed
a lineshape for electromagnetic absorption by assuming the
system may be thought of as independent Fukuyama-Lee-Rice (FLR)
domains, with randomness in the pinning frequencies of the 
individual domains satisfying $\Delta \omega_{pin}/\omega_{pin} \sim 1$.
By noting that there must be a length scale over which the 
variations in domain pinning frequency effectively decouple
the domains, an estimate for the expected lineshape
was found.

The model of Ref. \onlinecite{normand92} is in fact not very
different than ours.  One could think of the individual electrons
in our calculations as FLR domains, and the model investigated
would basically be the same.  
However, in that work it was
assumed that the length scale over which Coulomb coupling
may be ignored is the same as the FLR length, leading to
a resonance with $Q \sim 1$.  
This reasoning works well in the absence of a magnetic
field, because each domain has two possible polarizations
for their motion in a collective mode.  For the lowest
frequency collective modes, the domains can execute a
transverse motion that avoids long-range density fluctuations,
which are very high in energy.
Such long-range density fluctuations do appear when the
motion of the domains is longitudinal, and so such modes
contribute to the phonon density of states at high frequencies.
For the transverse modes, the coupling of motion among the
domains is weak, and the approximations of Ref. \onlinecite{normand92}
make sense.

In the presence of a magnetic field, however, it is not possible
to separate modes into transverse and longitudinal; these modes
are inevitably mixed.  The magnetic field causes the
domains to move in a circular fashion, essentially circulating
around their effective potential wells.  However, if different
domains circulate at different frequencies, there will necessarily
be long-range density fluctuations, so that such modes will be
high in frequency.  Low-frequency modes can be achieved if the
coupling between domains is explicitly included, so that correlations
in the motions of different domains may be introduced.  This means
in a magnetic field the system will have a {\it dynamical} correlation
length that is different than the {\it static} 
(FLR) correlation length,
and in this work we have seen that this dynamical correlation length
is extremely large.  We emphasize that this behavior is very
different than what occurs in most CDW systems, where the
static and dynamic correlation lengths are basically 
the same\cite{normand92,chitra}.
The new physics arises because of the unique combination of two-dimensions,
the long-range Coulomb interaction, and a magnetic field,
and leads to the sharp resonance found in this work.

Experimentally, it is somewhat surprising that only the most
recent measurements have uncovered the sharpness of this
resonance, whereas there have been a number of earlier
measurements of rf, surface acoustic wave, 
and microwave response in this 
system\cite{andrei88,paalanen92,willett94,glattli90}
which found broad resonances.  There may be several
reasons for this.  Firstly, in order to couple to the
2DEG, the experimental methods probe the system at a
finite $\vec{k}$, not with a purely spatially uniform
electric field.  From a practical viewpoint this is
necessary, as the oscillator strength for transitions
at small $\vec{k}$ is very small, making detection
of the absorbed energy difficult.  However, a repetition
of the analysis of Section \ref{sec:spin_rep} shows
that the absorption is sharp {\it only} for
$k=0$: the second-order correction in perturbation
theory diverges more strongly at finite $k$ than for
$k=0$, indicating that the response broadens as $k$ increases.
This observation is supported by the results of 
Ref. \onlinecite{mellor97}, for which several
absorption peaks are observed, presumably
representing harmonics of the fundamental
probing wavelength.  The widths of these
peaks are found to increase with increasing
frequency.  Thus, it may be that some of the
methods used in previous work coupled to or
mixed in large enough values of $k$ to wash
out the resonance.

A second important aspect of the result is that
it applies only at zero temperature; clearly one
expects thermal broadening of the resonance, which
we have not addressed in this work.  This means
that one can only expect such sharp resonances
at the lowest temperatures.  One study of the
temperature dependence of the resonance\cite{mellor97}
indicates that it is only apparent below 100mK, and
that it sharpens very rapidly in the range
50mK $\rightarrow$ 30mK.  It seems quite reasonable
that thermal broadening is responsible
for the absence of this sharp resonance in previous
experiments, especially since the temperature below
which the sharp resonance settles in may be sample-specific.

A number of open questions remain unresolved by this
work.  Firstly, because this is a zero-temperature
study, we have not been able to understand the
detailed absorption lineshape of Ref. \onlinecite{li97,mellor97}.
Beyond thermal effects, the large scale coherence
of the electron motion at zero temperature indicates
that specific dissipation mechanisms -- in particular,
edge states\cite{engel97,cote93} -- of the WC may prove important
in this context.  Beyond the lineshape, some aspects
of the magnetic field dependence of the resonance
remain unexplained: at the highest magnetic fields,
the resonance shows little or no field-dependence,
while the $Q$ of the resonance continues to increase
with field.  This work gives a natural explanation
for how the frequency may be an increasing function
of field in an interface pinning model; however, it
is not clear how one could obtain a field-independent
resonance\cite{com7}.  Investigations of these issues
are currently being pursued.

\section{Summary}

In this work, we studied the response  of a two-dimensional
Wigner crystal in a strong magnetic field to a spatially
uniform, time-dependent electric field, at zero temperature,
pinned by a disorder potential.  
A new approach to computing the response functions
of a localized electron system in the lowest Landau
level was introduced, the quantum harmonic approximation.
It was found that the 
response is sharp; i.e., there is a resonance that is
{\it not} disorder broadened.  For weak disorder, this
effect was shown within perturbation theory to survive
for macroscopically large samples, because of the
emergence of an extremely large
length scale $L_c$ that represents the
distance over which electrons oscillate together
in the lowest excited state of the system.
The fact that $L_c$ is so very large was shown
to result primarily from the long-range nature
of the Coulomb interaction.  A model of interface
pinning was shown to reproduce both the magnitude
and some aspects of the field dependence of the
resonance as observed in experiment.

\acknowledgements

The author is indebted to Prof. Sankar Das Sarma for 
telling him of the results of Ref. \onlinecite{li97},
as well as for many useful discussions and encouragement
through the course of this work.  Lloyd Engel, Chichun Li, 
Chris Mellor, and Dan Tsui are gratefully acknowledged
for making the results of their work available prior
to publication.  The author also thanks Steve Girvin
for a useful discussion.  This work was supported by
NSF Grant No. DMR95-03814, and the Research Corporation.

\smallskip

\noindent { \it Note added:} After the completion of this
work, the author became aware of a recent investigation\cite{chitra}
of the pinning properties of the magnetically-induced WC.
These authors are also able to explain the increase of
the pinning frequency with increasing field observed in
experiment.  However, the work does not find the unbroadened
resonance at zero temperature that is the focus of the
present paper.

\appendix

\section{Computation of Interaction Matrix Elements}

The interaction matrix elements $U_{m_1m_2m_3m_4}^{ij}$ can be
greatly simplified by the use of the strict translational
periodicity assumed in this work.  In general, they may
be written in the form
$$
U_{m_1m_2m_3m_4}^{ij}=
\int d^2r_1 d^2r_2 \int d^2q v(q) e^{-i\vec{q} \cdot 
(\vec{r_1}-\vec{r_2})}
$$
$$
\times 
\phi_{m_1}^*(\vec{r}-\vec{R}_i)\phi_{m_2}(\vec{r}-\vec{R}_i)
\phi_{m_3}^*(\vec{r}-\vec{R}_j)\phi_{m_4}(\vec{r}-\vec{R}_j),
$$
where
$\phi_{m}(\vec{r})$ is the $mth$ angular momentum state
centered around the origin.  The quantity $v(q)$ is the Fourier
transform of the electron-electron interaction, which
for most of this work will take the form $2\pi e^2/q$.
(To describe a screened Coulomb potential, one may take
$v(q)=2\pi e^2/\sqrt{q^2+q_c^2}$.  This was the form used
in the calculations leading to Fig. 2.)
Because we are imposing periodic boundary conditions,
the interaction needs to be replaced with one that
is periodic in a superlattice of unit cells, each with
$N$ electrons, whose positions inside each cell are
identical.  This is accomplished by making the
replacements
$\vec{q} \rightarrow \vec{G},\quad
\int d^2q \rightarrow {1 \over {v_c}}\sum_{\vec{G}}$
in the above expression, where $\lbrace \vec{G} \rbrace$ 
are the reciprocal lattice vectors of the superlattice,
and $v_c$ is the supercell area.
Making this replacement, and switching over to bra-ket notation,
we have
$$
U_{m_1m_2m_3m_4}^{ij} =
{1 \over {v_c}}\sum_{\vec{G}} v(\vec{G}) 
$$
\begin{equation}
\times 
<m_1|e^{-i\vec{G} \cdot \vec{r}}|m_2><m_3|e^{i\vec{G} \cdot \vec{r}}|m_4>
e^{-i\vec{G} \cdot (\vec{R}_i-\vec{R}_j) }.
\label{computed_mat_el}
\end{equation}
Although for the unscreened Coulomb interaction $v(\vec{G})$
diverges for $\vec{G}=0$, we may formally set it to zero
if one includes the effects of a uniform neutralizing background.
The matrix elements $<m_1|e^{-i\vec{G} \cdot \vec{r}}|m_2>$
may be computed with some work analytically. The result is
$$
<m_1|e^{-i\vec{G} \cdot \vec{r}} |m_2> =
\biggl[{{m_2!} \over {m_1!}} \biggr]
\biggl[{{(Gl_0)^2} \over {2}}\biggr]^{(m_1-m_2)/2}
$$
\begin{equation}
\times
L_{m_2}^{m_1-m_2}\biggl( { {(Gl_0)^2} \over {2} }\biggr)
e^{i(m_2-m_1)(\theta_{\vec{G}}+{{\pi} \over 2})-(Gl_0)^2/2},
\label{bra_ket}
\end{equation}
for $m_1\ge m_2$, where $\theta_G$ is the angle between the
the vector $\vec{G}$ and the $\hat{x}$ axis, and $L_{m_2}^{m_1-m_2}$
is an associated Laguerre polynomial.  The expression for
$m_2 > m_1$ may be obtained using 
$<m_1|e^{-i\vec{G} \cdot \vec{r}} |m_2> =
<m_2|e^{-i(-\vec{G}) \cdot \vec{r}} |m_1>^*$.
It is convenient to write this result in the form
$$
<m_1|e^{-i\vec{G} \cdot \vec{r}} |m_2> =
e^{i(m_2-m_1)(\theta_{\vec{G}}+{{\pi} \over 2})} F_{m_1m_2}(G),
$$
with
$$
F_{m_1m_2}(G)= 
\biggl[{{m_2!} \over {m_1!}} \biggr]
\biggl[{{(Gl_0)^2} \over {2}}\biggr]^{(m_1-m_2)/2}
$$
$$
\times
L_{m_2}^{m_1-m_2}\biggl( { {(Gl_0)^2} \over {2} } \biggr) e^{-Gl_0^2/2}
$$
for $m_1 \ge m_2$.  For $m_1 < m_2$, the expression for $F_{m_1m_2}(G)$
has the same form as above, only with the indices $m_1,~m_2$
interchanged.

The interaction matrix element can now be written in the form
$$
U_{m_1m_2m_3m_4}^{ij} =
{1 \over {v_c}}\sum_{\vec{G}} v(\vec{G}) 
e^{-i\vec{G} \cdot (\vec{R}_i-\vec{R}_j)}
F_{m_1m_2}(G)F_{m_3m_4}(G)
$$
$$
\times
(-1)^{m_4-m_3}i^{|m_1-m_2|+|m_3-m_4|}
e^{i[m_2-m_1+m_4-m_3]\theta_{\vec{G}}}
$$
which is particularly convenient for calculations.
For a given value of $\vec{R}_i-\vec{R}_j$
one in practice can include a very large
number of reciprocal lattice vectors 
in the sum; for example, our calculations
with $N=1024$ electrons include approximately
60,000 different values of $\vec{G}$ in the
sum.  Because we include so many of these,
the QHA can accurately reflect what happens
at length scales shorter than a magnetic
length, allowing one to treat disorder
potentials that vary on a short length
scale.  This is the advantage the QHA has
over the method of Ref. \onlinecite{cote91}, which
is typically limited to several hundred
values of $\vec{G}$.

In the text, we took advantage of the sum
rule
\begin{equation}
\sum_l
U_{m_1m_200}^{\prime il} \equiv
\sum_l U_{m_1m_200}^{il} - ~~U_{m_1m_200}^{ii} \propto \delta_{m_1m_2}
\label{sumruleapp}
\end{equation}
which we now demonstrate.  We begin with simple observation that
$$\sum_j e^{i\vec{G}\cdot \vec{R}_j} = N \sum_{\vec{g}}
\delta_{\vec{g},\vec{G}},$$
where the $\lbrace\vec{g}\rbrace$ are the reciprocal lattice 
vectors of the electron lattice (i.e., not the superlattice).
Then
$$
\sum_l U_{m_1m_200}^{il}=
N\sum_{\vec{g}} v(g)<m_1|e^{-i\vec{g}\cdot\vec{r}}|m_2>
<0|e^{i\vec{g}\cdot\vec{r}}|0>.
$$
From Eq. \ref{bra_ket} it is clear that $<0|e^{i\vec{g}\cdot\vec{r}}|0>$
depends only on the magnitude of $\vec{g}$ and not its orientation,
so that the orientation angle enters the sum above through
$<m_1|e^{-i\vec{g}\cdot\vec{r}}|m_2> \propto e^{i(m_2-m_1)\theta_{\vec{g}}}$.
Since each reciprocal lattice vector has five others of the
same magnitude oriented at angles that are integral multiples
of $\pi/3$ away from $\theta_{\vec{g}}$, if follows that
the phase factor will cause the above sum to vanish unless
$m_1=m_2$.  Thus
\begin{equation}
\sum_l U_{m_1m_200}^{il} \propto \delta_{m_1m_2}.
\label{sumrule_a}
\end{equation}
Similarly, for $U_{m_1m_200}^{ii}$ we have
$$U_{m_1m_200}^{ii}=\quad\quad\quad\quad$$
$$
\sum_{\vec{G}} v(G)<m_1|e^{-i\vec{G}\cdot\vec{r}}|m_2>
<0|e^{i\vec{G}\cdot\vec{r}}|0>.
$$
As above, the quantity entering the sum involves the orientation
of $\vec{G}$ only through $e^{i(m_2-m_1)\theta_{\vec{G}}}$,
so this sum must also vanish unless $m_1=m_2$.  Together
with Eq. \ref{sumrule_a}, this proves the sum rule
Eq. \ref{sumruleapp}.

\onecolumn

\begin{figure}
 \hbox { \epsfysize=8.in
 \epsffile{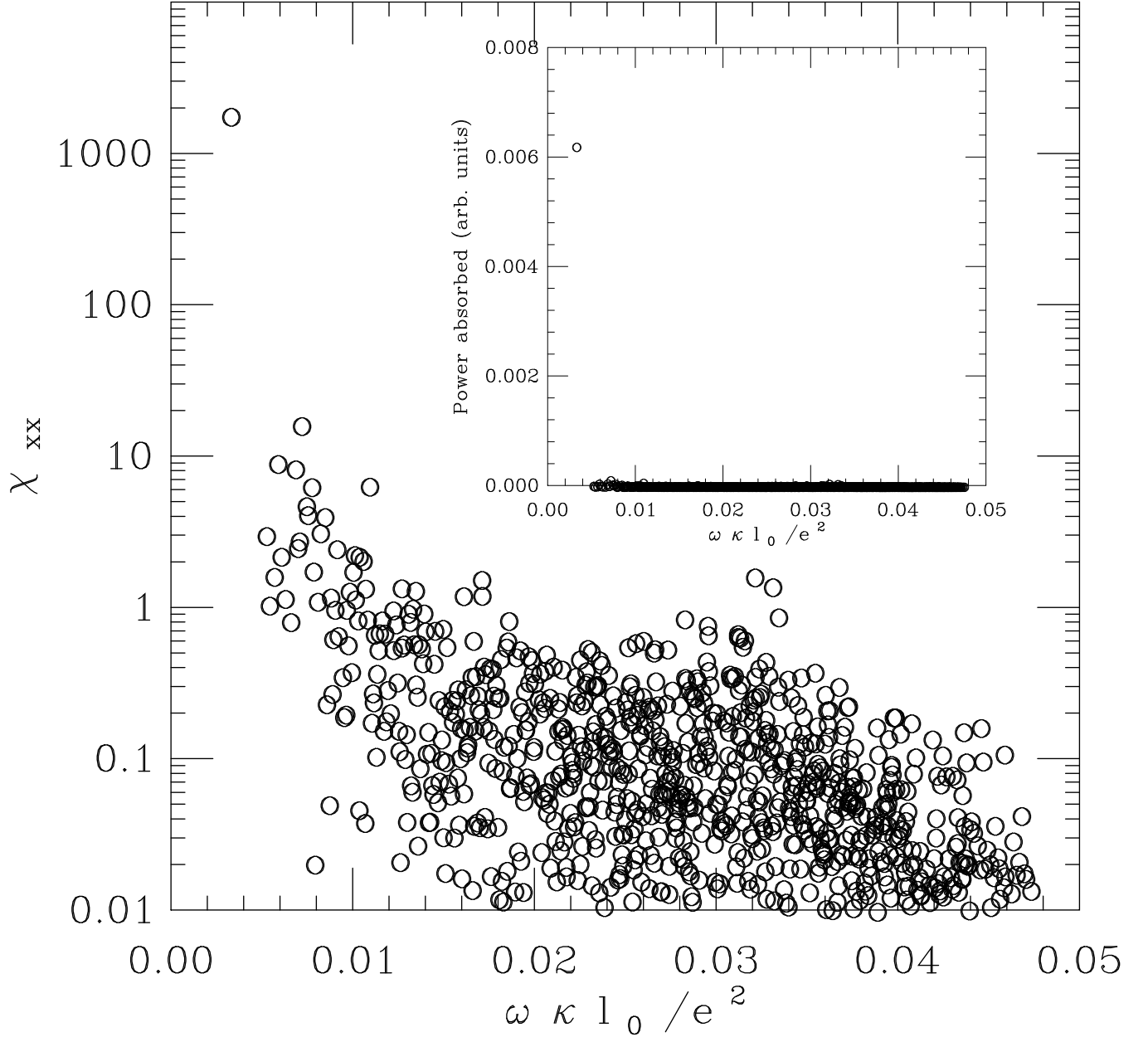}  }
\caption{
Center of mass response function $\chi^{xx}_{CM}$ for a system of
$N=1024$ electrons at $\nu=0.2$.  Half the sites are chosen at
random to have a pinning potential $\Delta U_i = 0.01e^2/\kappa l_0$
(see text);
the other half are unpinned.  
Two states per site were retained in this calculation (see text).
Inset: Relative power absorption
for the same system.
}
\end{figure}
\vfill\eject
\begin{figure}
 \hbox {  \includegraphics{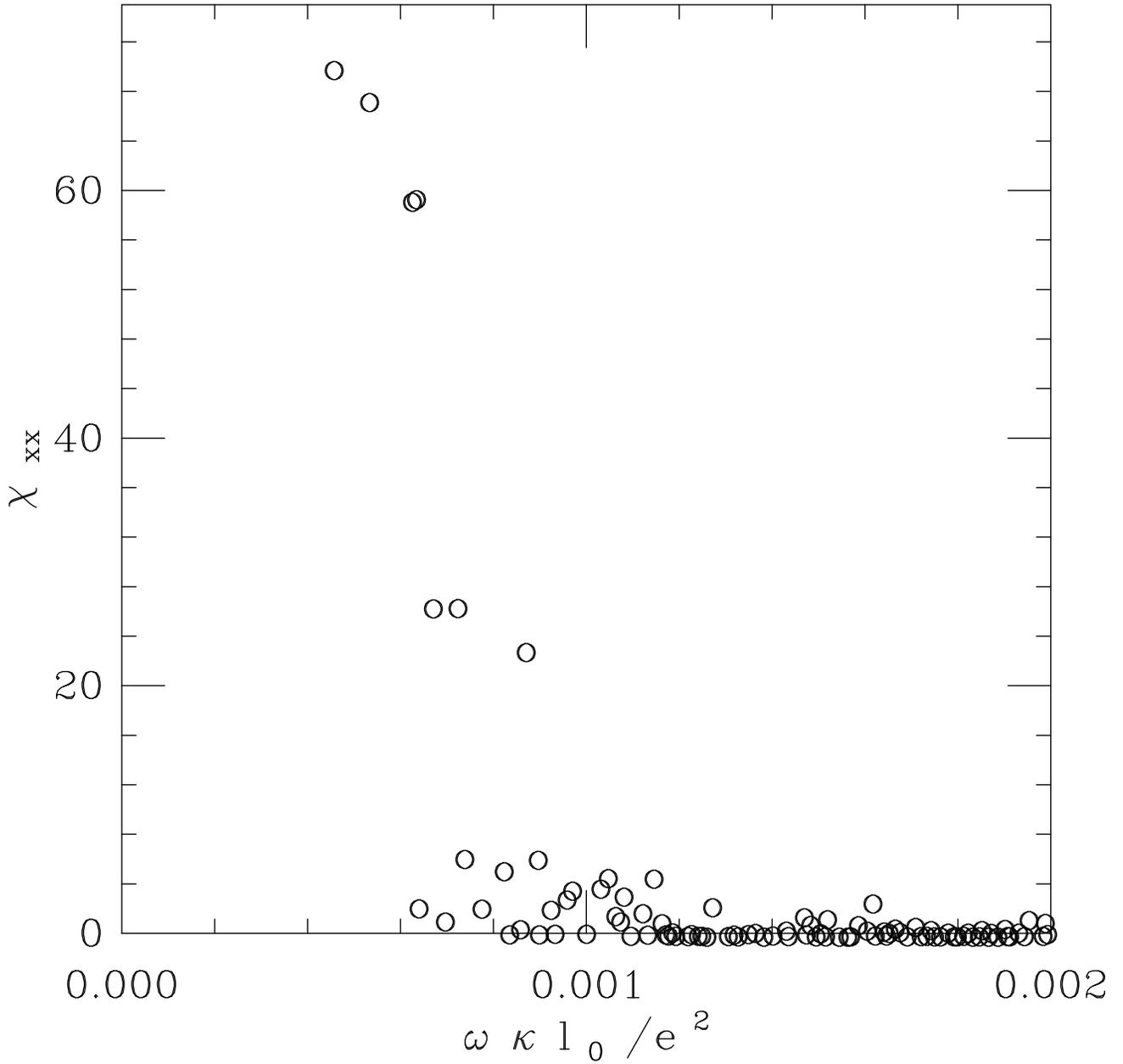}}
 \caption{
 Center of mass response function $\chi^{xx}_{CM}$ for a system
 of size $N=225$ electrons at $\nu=0.2$
 with a screened Coulomb potential of the form
 $v(q) = 2\pi e^2/ \kappa (q^2+q_c^2)^{1/2}$, 
 with $q_c=2.0/l_0$.
 Half the sites are chosen at
 random to have a pinning potential $\Delta U_i = 0.01e^2/\kappa l_0$
 (see text); the other half are unpinned.  Two states per site
 were retained in this calculation.  The result indicates that
 screening broadens the response, so that the sharpness of the response
 illustrated in Fig. 1 is related to the long-range nature of
 the Coulomb interaction.
 }
 \end{figure}
 
 \vfill\eject
 
\begin{figure}
 \hbox {  \includegraphics{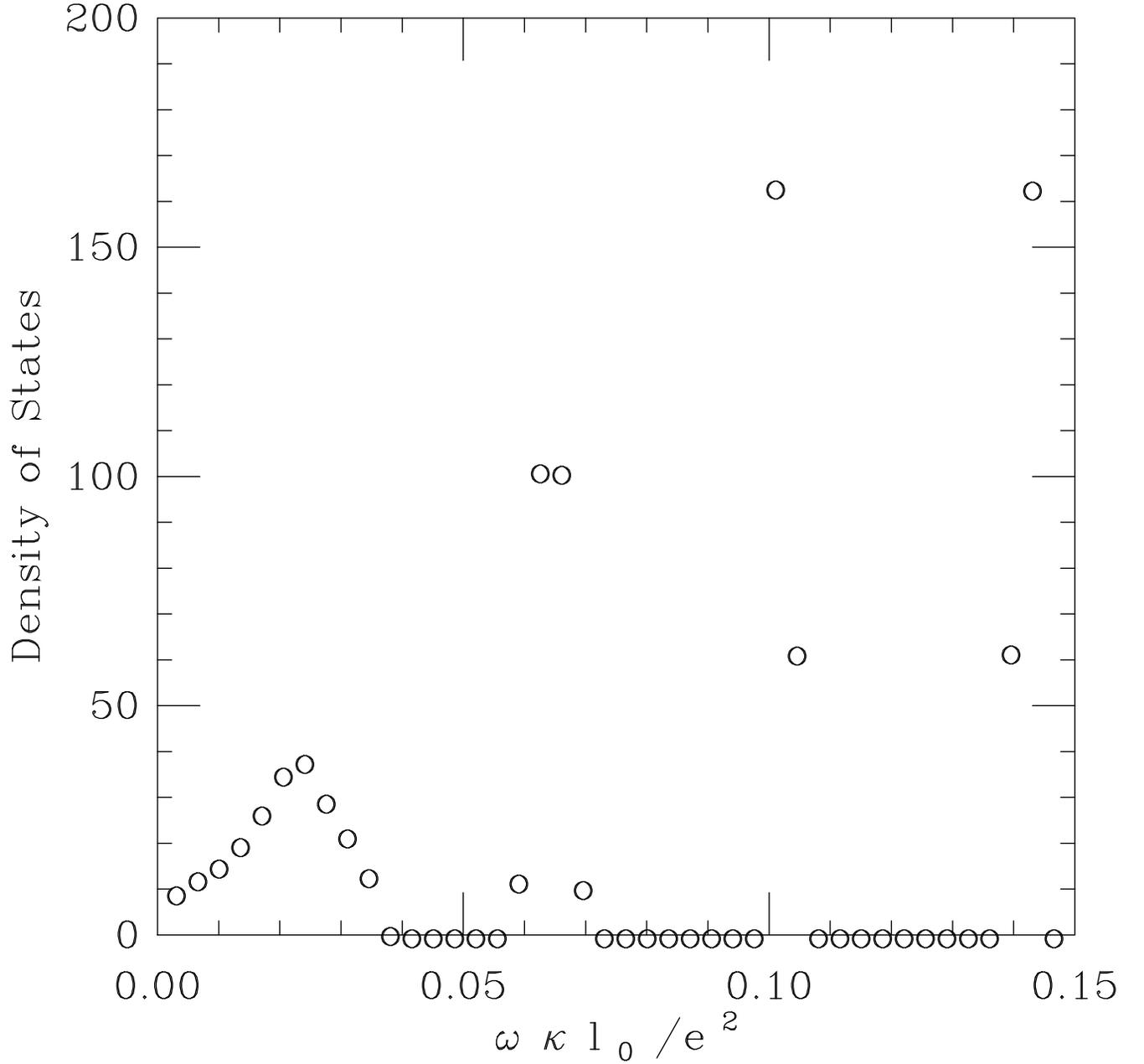}}
 \caption{
Collective mode density of states for an unpinned WC as calculated using the
quantum harmonic approximation.  $N=225$, $\nu=0.2$, and 5 states
per site are retained in the calculation.  The result illustrates
that including higher order angular momentum states introduces
collective modes at high energy, which may be interpreted as
an analog of Wannier excitons (see text).  Such collective modes
have little effect on the low energy dynamics of the system, so
that one may drop the high angular momentum states without introducing
serious errors in the pinning response of the system.
 }
 \end{figure}
 
\vfill\eject

\begin{figure}
 \vbox {  \includegraphics{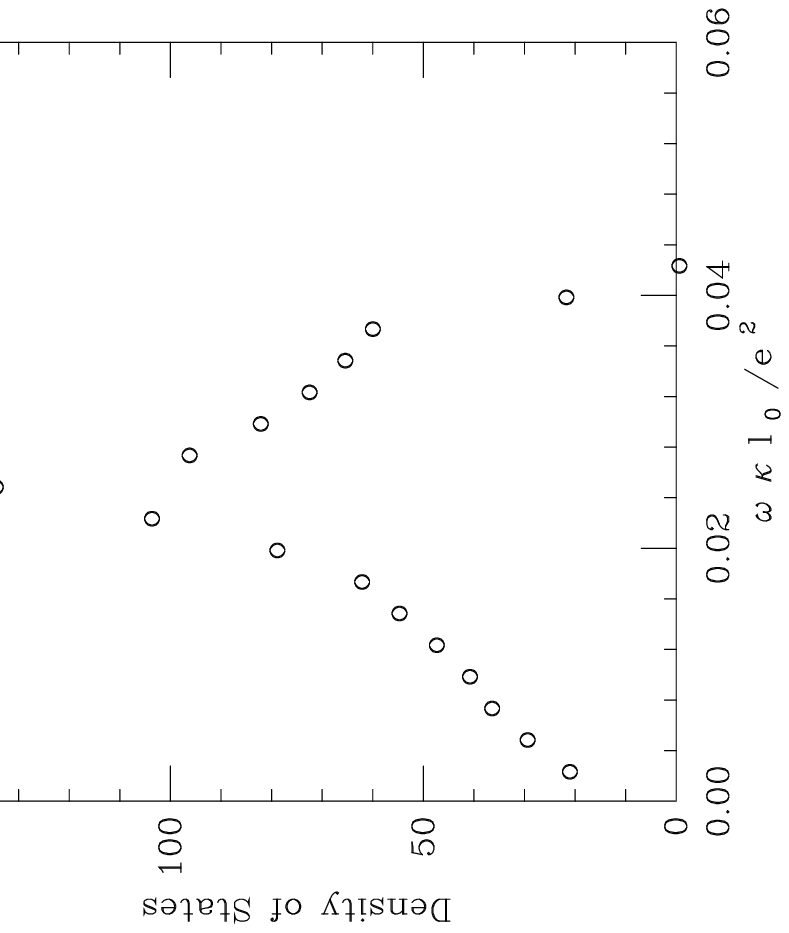}}
 \caption{
Collective mode density of states for an unpinned WC as calculated using the
quantum harmonic approximation (QHA).  $N=1024$, $\nu=0.2$, and 2 states
per site are retained in the calculation.  This has a form consistent
with the phonon density of states for a classical, unpinned WC in
a strong magnetic field, and demonstrates that the QHA produces
sensible results for the test case of a WC in the absence of
disorder.
 }
 \end{figure}
\vskip 11.5cm 
\begin{figure}
 \vbox {  \includegraphics{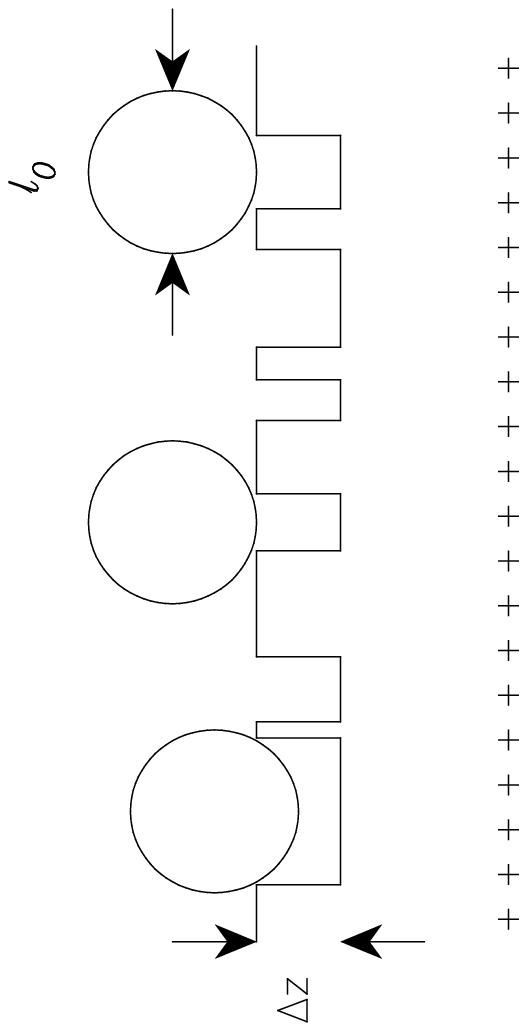}}
 \caption{
Schematic representation of the interface disorder model.  The interface
has pits and terraces so that the setback from the ionized donor plane
varies by a distance of order $\Delta z$.  Electrons localized by the magnetic
field into wavepackets in the $x-y$ plane 
of size scale $\sim l_0$ may lower their
energy by moving their centers close to pit centers, since 
in the area of the pit the
electron will be closer to the donor layer.  For larger pits
or decreasing magnetic length, more of the electron wavepacket
overlaps with the pit region, decreasing
the energy of the wavepacket.
  }
 \end{figure} 
 \vfill\eject

\vfill\eject

\begin{figure}
 \hbox {  \includegraphics{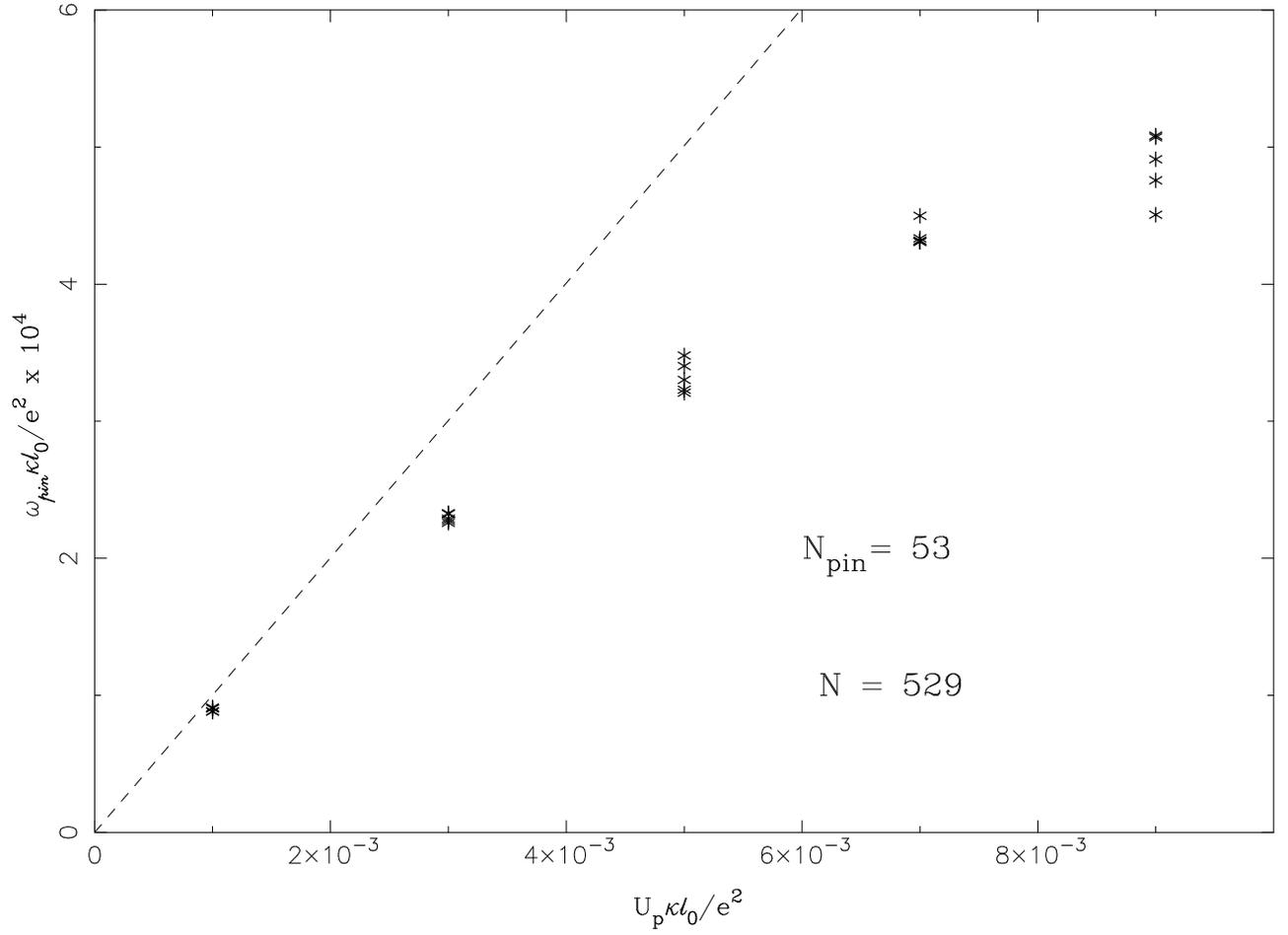}}
 \caption{
Pinning frequency $\omega_{pin}$ for a system of $N=529$ electrons,
at filling fraction $\nu=0.2$, as a function of the pit potential.
10\% of the electrons are pinned at random sites with energy 
$\Delta U=U_p$; the remaining electrons are unpinned.
Stars are the results
of the QHA for five different disorder realizations at each value
of $U_p$, with two states per site retained in each calculation.  
Dotted line is the expected pinning frequency
as calculated in lowest order perturbation theory.  For small
values of $U_p$ the approaches agree quite well; for larger
values the perturbation theory
sets an upper bound on the pinning frequency.
 }
 \end{figure} 
 
 \vfill\eject
 
\begin{figure}
 \hbox {  \includegraphics{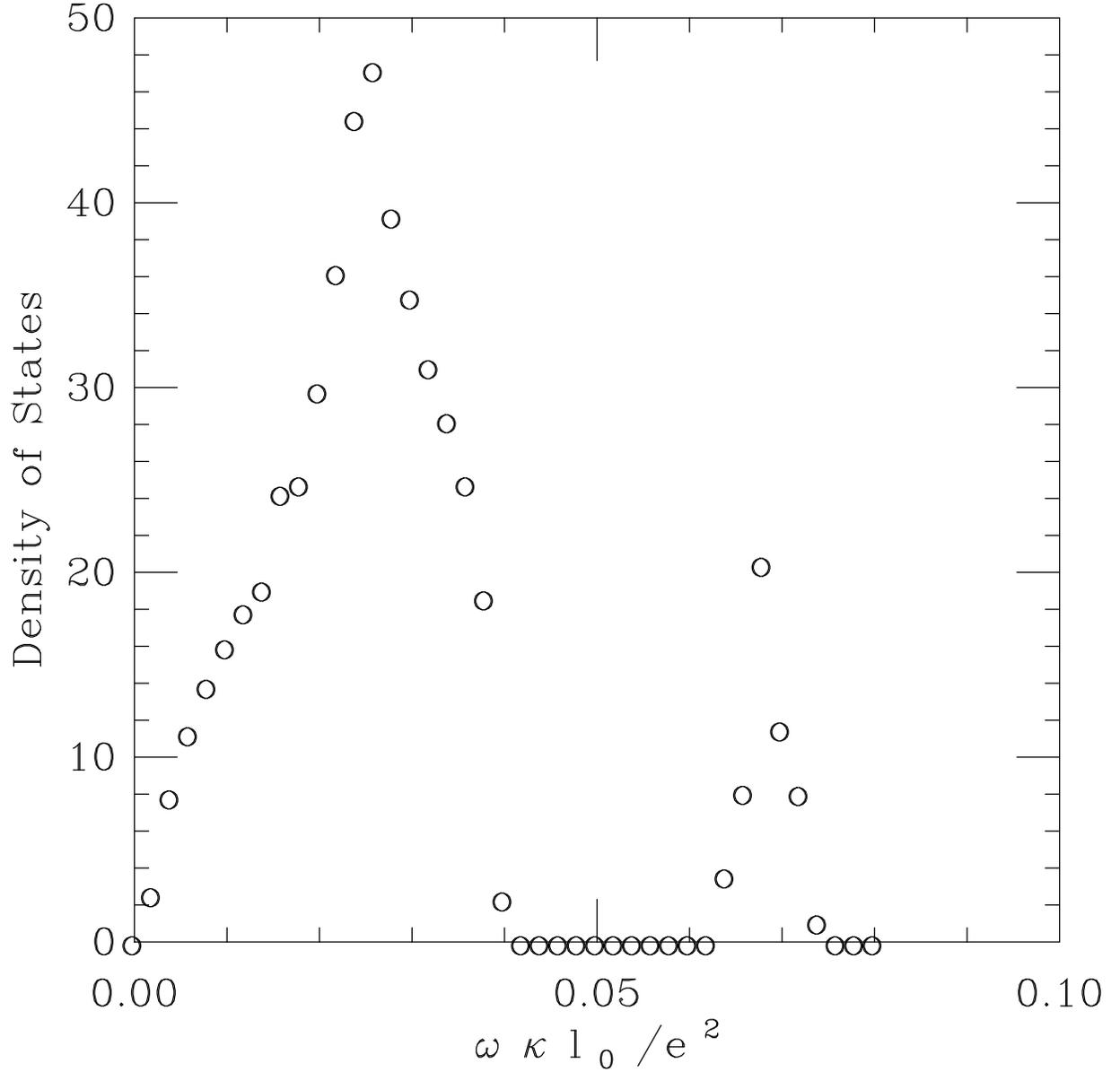}}
 \caption{
Density of states 
for a strongly pinned WC, 
computed using the QHA,
with $N=529$, $\nu=0.2$, and two states retained per site.
10\% of sites have a pinning potential 
$v_0+\Delta U_i=U_p=0.04e^2/\kappa l_0$,
a relatively large pinning potential; the rest
are unpinned.  A high energy set of
collective modes may clearly be seen separated above the 
main peak.  These modes arise due to localized collective modes
of the strongly pinned electrons.  For the modes in the lower, main peak,
the pinned electrons are essentially stationary.
 }
 \end{figure} 
 
\vfill\eject
 
\begin{figure}
 \hbox {  \includegraphics{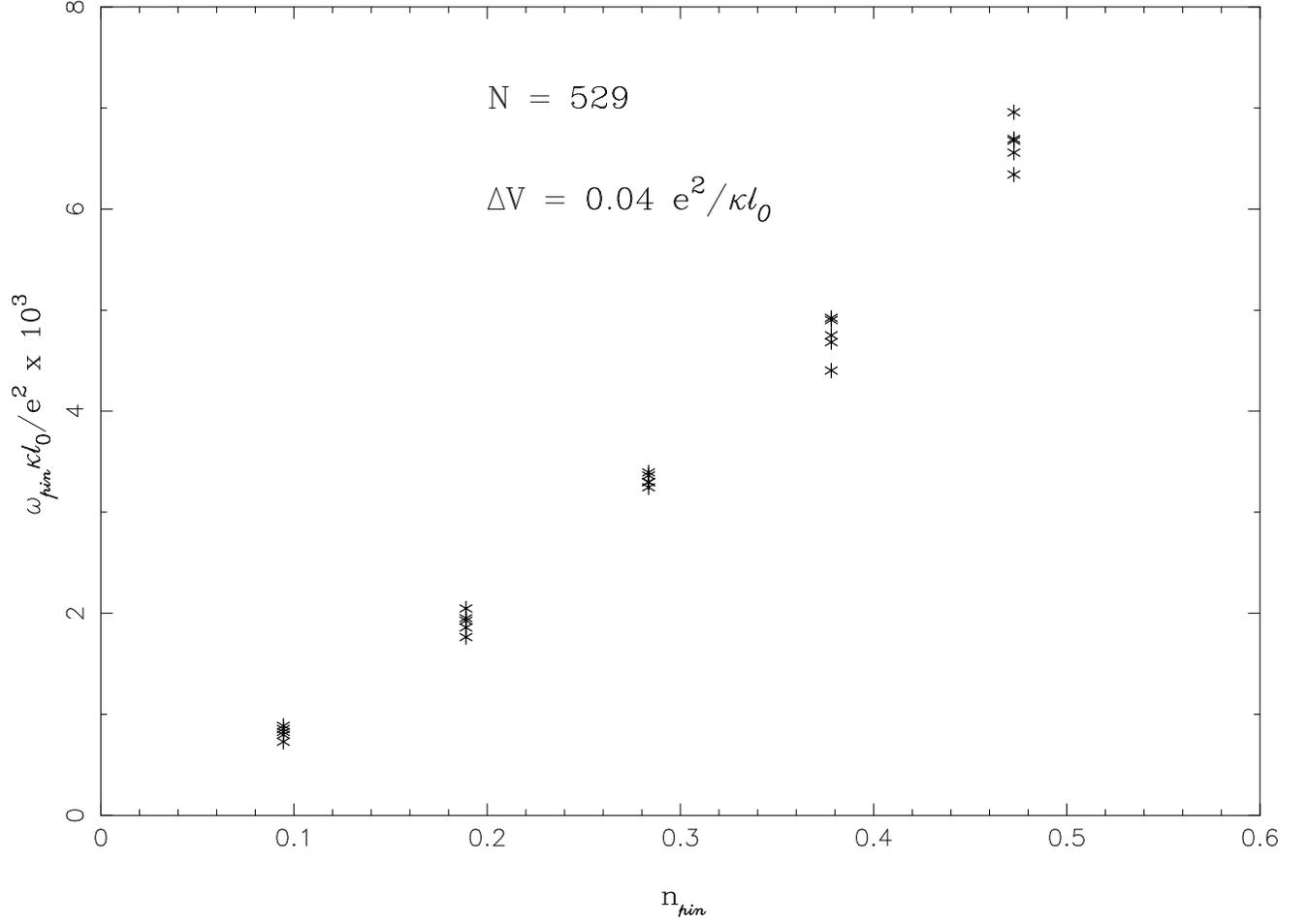}}
 \caption{
Pinning frequency $\omega_{pin}$ for a system of $N=529$ electrons
with strong pinning centers.  $\omega_{pin}$ was
computed using the QHA with two states per site
at filling fraction $\nu=0.2$ as a function of $n_{pin}$,
the fraction of pinned sites, with 
$v_0+\Delta U_i=U_p=0.04e^2/\kappa l_0$ the potential of the
pinned sites.  ($1-n_{pin}$ of the sites are unpinned.)
Stars are the results
for five different disorder realizations at each value
of $n_{pin}$.  The computed frequencies are significantly lower
than what is expected from an estimate based on strong pinning
of an elastic medium.
 }
 \end{figure}
 
\vfill\eject

\begin{figure}
 \hbox {  \includegraphics{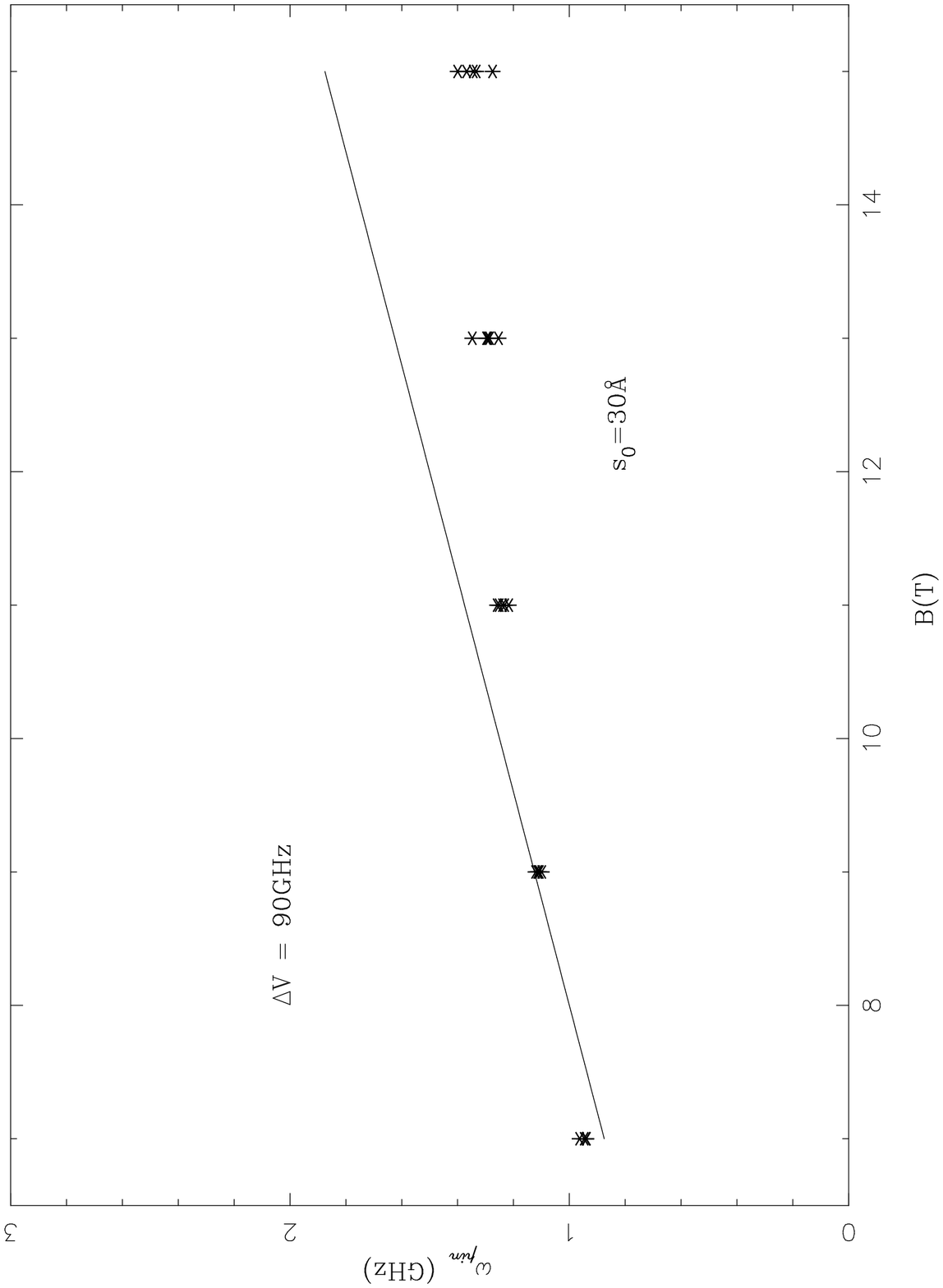}}
 \caption{Pinning frequency $\omega_{pin}$ 
in the interface pinning model
as a function of magnetic field,
for $\Delta V = 90$GHz corresponding to a pit depth
of $\Delta z = 10\AA$. The average pit size
$s_0= 30$\AA, and
the electron density in this
figure is fixed at
$\rho_0=5 \times 10^{10}$cm$^{-2}$.
Solid line illustrates the perturbative estimate
of $\omega_{pin}$ for a pit density of $n_i = 2.5 \times 10^{11}cm^{-2}$.
The magnitude of the
pinning frequency may be seen to increase linearly with
magnetic field, due to the decreasing magnetic length
of the Gaussian orbitals in which the electrons reside
in the groundstate.  Stars represent the results
of a quantum harmonic approximation (QHA) calculation
with the same parameters except for a slightly increased
pit density,
$n_i = 3.0 \times 10^{11}cm^{-2}$.  Results for five different
disorder realizations are shown for $B=7,~9,~11,~13$ and 15$T$.
 }
 \end{figure}


\begin{references}
\bibitem{ferccom} The microscopic model adopted in this work is qualitatively
similar to that of Ref. \onlinecite{ferconi93}.  That work however focused
on the value of the pinning frequency and did not directly
compute the electromagnetic response of the WC.  We believe a
computation of this quantity within the model of 
Ref. \onlinecite{ferconi93} would lead to a sharp
density response, as found in this work.
\bibitem{com0} Once we have projected into the lowest Landau level,
there is no distinction between 
the cases when the charged particles in the WC are
electrons and when they are holes.
For clarity we refer to them throughout the text as electrons,
even though in many experiments (particularly Refs. \onlinecite{li97}
and \onlinecite{mellor97}) the carriers are in fact holes.
\bibitem{com1} It should be noted that although pinning 
via charged impurities as
discussed in Ref. \onlinecite{ruzin92} is inconsistent with 
experimentally measured pinning frequencies, it may well be
true that this mechanism sets the energy scale for the
static electric field threshold found in experiments.
The possibility that non-linear $IV$ characteristics and
low frequency ac response are dominated by different
pinning mechanisms could explain the difficulties
in reconciling the experimental measurements of these with
a model that has a reasonable crystalline correlation
length\cite{normand92,ferconi93}.
\bibitem{com2} Note this is where the approximation of ignoring overlaps
between states localized at different sites is introduced.
\bibitem{com3} One may be more careful and directly compute 
$\int d^2r <\vec{j}(\vec{r},\omega)> \cdot <\vec{E}_{ind}(\vec{r},\omega)>$
to account for spatial correlations between the current and 
induced electric field for a given disorder realization.  However,
for the very sharp responses we find at zero temperature,
this difference is purely academic.  At finite temperature
where the response is broadened, this latter approach is
preferable, although it is likely that the long-range nature
of the Coulomb interaction will make the spatial variations
in $<\vec{E}_{ind}>$ quite weak.
\bibitem{com4} As usual, when representing spin waves in terms of
bosonic degrees of freedom, one must keep in mind that the
number of bosons per site is constrained to be no more than
1.  Provided one is interested in states with very few
spin waves (as is the case here), the constraint does not
need to be enforced explicitly.
\bibitem{com5} These properties are easily understood if one recognizes
that the matrix is precisely the same as the one
that must be diagonalized to express the pseudospin-wave creation and
annihilation operators in terms of the site boson operators
$b_i,~b_i^{\dag}$.
\bibitem{com6} We expect for extremely large densities of pinned
sites that one should observe a broadening of the absorption.  However,
the density required for this is too large to be of physical
interest.
\end{references}
\end{document}